\documentclass[times,twocolumn,final]{elsarticle}
\usepackage{medima}
\usepackage{framed,multirow}

\usepackage{amssymb}
\usepackage{latexsym}

\usepackage{url}
\usepackage{xcolor}
\usepackage{algorithmic}
\usepackage{graphicx}
\usepackage{textcomp}

\usepackage{multirow}
\usepackage{caption}
\usepackage{subcaption}

\usepackage{physics}
\usepackage{booktabs}
\usepackage{comment}
\usepackage{adjustbox}
\usepackage{xcolor}

\usepackage[shortlabels]{enumitem}
\usepackage{makecell}
\usepackage{array}

\usepackage{color}
\usepackage{comment}
\usepackage{multirow}

\usepackage[hidelinks]{hyperref}

\definecolor{newcolor}{rgb}{.8,.349,.1}

\journal{Computational and Structural Biotechnology Journal}

\begin{document}

\verso{W. Zhang, S. Yang, M. Luo\textit{et~al.}}

\begin{frontmatter}

\title{Keep It Accurate and Robust: An Enhanced Nuclei Analysis Framework}%

\author[1]{Wenhua Zhang}
\author[2]{Sen Yang}
\author[3]{Meiwei Luo}
\author[4]{Chuan He}
\author[2]{Yuchen Li\corref{cor1}}
\author[3]{Jun Zhang\corref{cor1}}
\author[2]{Xiyue Wang\corref{cor1}}
\author[5]{Fang Wang\corref{cor1}}

\cortext[cor1]{Corresponding author. E-mail addresses: 578563666@qq.com(F. Wang), ycli16@stanford.edu(Y. Li), xdzhangjun@gmail.com(J. Zhang), xiyuew@stanford.edu(X. Wang)}
  
\address[1]{Institute of Artificial Intelligence, Shanghai University, Shanghai 200444, China. (Email: winniezhangcoding@gmail.com)}
\address[2]{Department of Radiation Oncology, Stanford University School of Medicine, Stanford, CA 94305 USA. (Email: xiyuew@stanford.edu, sen.yang.scu@gmail.com, ycli16@stanford.edu)}
\address[3]{Tencent AI Lab, Shenzhen 518057, China. (Email: lmw99@foxmail.com, junejzhang@tencent.com)}
\address[4]{Shanghai Aitrox Technology Corporation Limited, Shanghai, China, 200444, China. (Email: hech@fosun.com)}
\address[5]{Department of Pathology, The Affiliated Yantai Yuhuangding Hospital of Qingdao University, Yantai, 264000, China. (Email: 578563666@qq.com)}

\received{19 January 2022}
\finalform{10 May 2013}
\accepted{13 May 2013}
\availableonline{15 May 2013}

\begin{abstract}
%%%
Accurate segmentation and classification of nuclei in histology images is critical but challenging due to nuclei heterogeneity, staining variations, and tissue complexity. {Existing methods often struggle with limited dataset variability, with patches extracted from similar whole slide images (WSI), making models prone to falling into local optima. Here we propose a new framework to address this limitation and enable robust nuclear analysis. Our method leverages dual-level ensemble modeling to overcome issues stemming from limited dataset variation. Intra-ensembling applies diverse transformations to individual samples, while inter-ensembling combines networks of different scales. We also introduce} enhancements to the HoVer-Net architecture, including updated encoders, nested dense decoding and model regularization strategy.
We achieve state-of-the-art results on public benchmarks, including 1st place for nuclear composition prediction and 3rd place for segmentation/classification in the 2022 Colon Nuclei Identification and Counting (CoNIC) Challenge. This success validates our approach for accurate histological nuclei analysis. Extensive experiments and ablation studies provide insights into optimal network design choices and training techniques. In conclusion, this work proposes an improved framework advancing the state-of-the-art in nuclei analysis. We release our code and models (\url{https://github.com/WinnieLaugh/CONIC_Pathology_AI}) to serve as a toolkit for the community.
%%%%
\end{abstract}

\begin{keyword}
\KWD digital pathology\sep deep learning\sep nuclei segmentation\sep nuclei classification
\end{keyword}

\end{frontmatter}

%\linenumbers

% main text
\section{Introduction}
\label{sec:introduction}
Advancing the analysis of pathology images remains {crucial} for improving cancer diagnosis and prognosis \citep{rubin2008clinicopathologic}. Pathologists heavily rely on information extracted from pathology images to determine tumor grades \citep{fleming2012colorectal}, predict patient survival rates \citep{alsubaie2018bottom}, and anticipate responses to various therapies \citep{srinidhi2020deep}. Among the many analytical steps involved, nuclear segmentation and classification, as well as cellular composition prediction, are particularly fundamental, as they enable the extraction of cell-based features useful for numerous downstream tasks \citep{alsubaie2018bottom,lu2018nuclear,sirinukunwattana2018novel,srinidhi2021deep}. While manual analysis is vulnerable to intra- and inter-observer variability \citep{elmore2015diagnostic}, automatic methods can enable consistent and rapid analysis of these tasks. 

Researchers have leveraged image processing techniques since the 1990s to analyze morphological cell features \citep{chan1996expert,haroske1996nuclear,comaniciu1999image}. 
The recent advent of deep learning has further accelerated the field, with neural networks becoming a popular choice for solving biomedical challenges \citep{jia2009mechanisms,zou2018predicting,lin2014libd3c,sedaghat2017clinical,shen2016evolving,chen2011support}. This has led to the development of several deep learning approaches for nuclear segmentation and classification \citep{graham2019hover,dawood2021albrt,sirinukunwattana2016locality,schmidt2018}, which have demonstrated substantial improvements over traditional techniques.

Instance segmentation in digital pathology is challenging due to the high degree of overlap between adjacent nuclei~\citep{irshad2013methods}. Recently, various methods have been proposed to tackle this challenge. These methods can be broadly categorized into two types: segmentation-only approaches and simultaneous segmentation and classification approaches. Segmentation-only approaches focus on segmenting individual instances without considering their classification information\citep{abdel2022efficient,hassan2021efficient}. On the other hand, simultaneous segmentation and classification approaches aim to segment and classify instances simultaneously. For example, methods based on Mask R-CNN \citep{bancher2021improving,prabhusegmentation} propose regions of interest before segmenting and classifying objects within. However, inaccurate region proposals can degrade segmentation and classification. An alternative approach is to use U-Net architectures \citep{graham2019hover,dogar2023attention,schmidt2018}, which directly predict pixel-level maps and apply post-processing for instance segmentation and classification. These methods are particularly well-suited for highly overlapping scenarios.

Cellular composition prediction can provide valuable insights into tumor types and cell populations \citep{he2021deeply,zhu2021real}. As manual counting is tedious and subjective, automated approaches typically {employ} either regression to directly predict cell counts~\citep{khan2016deep,xue2016cell,dawood2021albrt} or object detection to first identify cells and then count them~\citep{xing2013automatic,cirecsan2013mitosis}. While regression methods are simple and effective, detection provides both segmentation and counting within a single framework.

In this work, we present a novel deep-learning framework for joint nuclear segmentation, classification, and cellular composition prediction in digital pathology images. The framework is designed to improve robustness through ensemble modeling, fusing predictions from diverse improved base models. 

Our framework consists of specialized base models, each trained on different data folds with varying encoder backbones. To enhance computational efficiency, we replace the heavy decoder branches of prior arts \citep{graham2019hover}. The predictions from each base model are then fused via averaging and post-processing to harness diversity. For cellular composition, detected nuclei are counted by class, establishing interdependency between nuclear identification and cellular composition analysis.

To evaluate the performance of our framework, we participated in the 2022 CONIC Challenge and achieved top rankings in cellular composition prediction and 3rd in nuclear segmentation/classification. Additional experiments on external datasets PanNuke \citep{gamper2019pannuke,gamper2020pannuke} and MoNuSAC \citep{verma2021monusac2020} also demonstrate superior performance over the state-of-the-art.

We list our key contributions as follows:
\begin{itemize}
\item Proposing a dual-ensemble system leveraging intra-model and inter-model ensembling to avoid overfitting by a single model that may fall into local optima.
\item Developing an enhanced architecture for nuclear instance segmentation and classification by upgrading HoVer-Net with advanced encoders, transformed decoders, and regularization techniques.
\item Demonstrating state-of-the-art performance of the enhanced framework on publically benchmarks for nuclear instance segmentation, classification, and composition prediction.
\end{itemize}

\section{Related Work}

\begin{figure*}[t!]
\centering
\includegraphics[width=\textwidth]{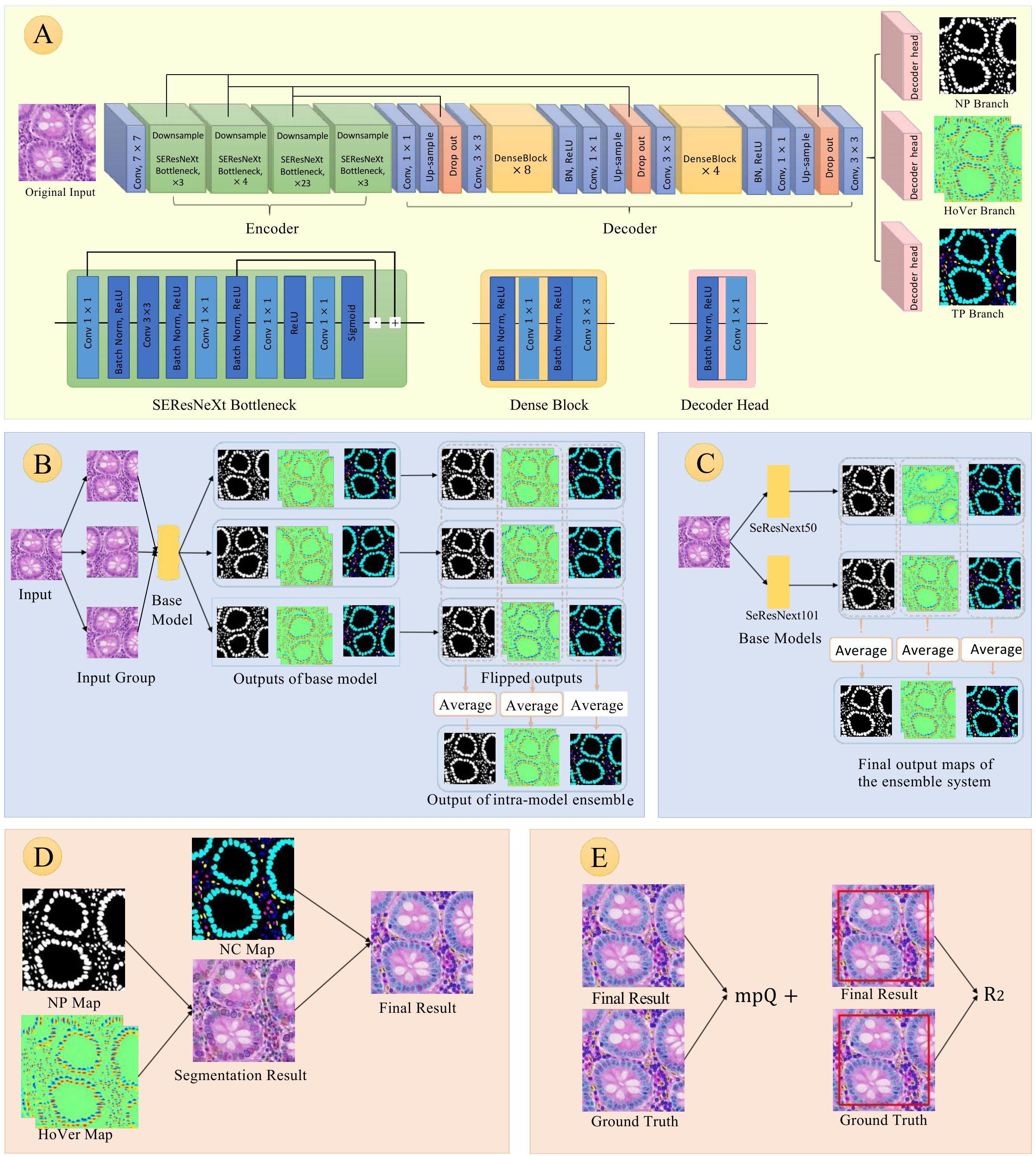}
\caption{Overview of the proposed method. A) The base model architecture utilizes an encoder-decoder structure adapted from HoVer-Net \citep{graham2019hover}. To improve compactness and effectiveness, the three decoder branches have been consolidated into one. Heavy dropout layers are also incorporated in the decoder to regularize training. While any encoder with a similar structure can be used, this example implements a SEResNeXt50 backbone \citep{hu2018squeeze}. B) The intra-model ensemble approach augments the input image with horizontal and vertical flips. Each base model makes predictions on the original and flipped inputs, which are averaged together after flipping the outputs back to the original orientation. This allows the network to leverage multiple views of the input during inference. C) The inter-model ensemble averages the output maps from base models with different encoder backbones to improve robustness. D) Post-processing utilizes the output maps for final instance recognition. Instances are first segmented using the nuclear presence (NP) and HoVer maps. These segmentation results are then grouped with the {nucleus classification (NC) map} for instance classification. E) Evaluation uses the multi-class panoptic quality ($mPQ+$) metric applied to the full input patch and multi-class coefficient of determination ($R^2$) applied to the $224\times224$ pixel center region.}
\label{fig:base_model}
\end{figure*}

\subsection{Nuclear Segmentation and Classification}
Nuclear segmentation and classification have gained significant attention in recent years within computational pathology research \citep{irshad2013methods,graham2019hover,yao2021pointnu,bancher2021improving,prabhusegmentation,abdel2022efficient,hassan2021efficient,schmidt2018}. The field can be broadly categorized into two types: segmentation-only approaches and simultaneous segmentation and classification approaches.

Segmentation-only approaches focus on segmenting individual instances without considering their classification information. For instance, Abdel-Nasser et al. proposed a staining-invariant encoder and a weighted hybrid dilated convolution block to efficiently segment nuclear instances without color normalization \citep{abdel2022efficient}. Similarly, Hassan et al. proposed a learnable aggregation network to ensemble a set of individual nuclear segmentation models \citep{hassan2021efficient}.

Simultaneous segmentation and classification approaches aim to segment and classify instances simultaneously. For example, methods based on Mark R-CNN propose candidate regions and then attempt to segment nuclear instances within each region proposal before classifying the detected nuclei \citep{bancher2021improving,prabhusegmentation}. However, the prevalence of overlapping nuclei introduces significant noise during region proposal, which cascades into errors in downstream segmentation and classification. To address this, some researchers have explored alternative frameworks based on U-Net architectures. A two-stage approach is often taken, where nuclei coordinates are first detected and then patches are cropped around these coordinates to feed into a separate classification model \citep{sirinukunwattana2016locality}. Other methods opt for a single model that elegantly handles both tasks of segmentation and classification simultaneously \citep{graham2019hover,yao2021pointnu,schmidt2018}. Notably, the advent of HoVer-Net \citep{graham2019hover} has a significant contribution in this area, which inspired several innovative variants. One such variant integrates tissue segmentation \citep{wang2023improved} to provide a contextual framework that aids in the precise identification of nuclei. Another variant introduces a multiple filter unit \citep{vo2023mulvernet}, designed to capture a wider range of features. In parallel, the CellViT \citep{horst2024cellvit} represents a shift towards utilizing vision transformers as encoders, taking advantage of the rich feature extraction capabilities of large-scale pretrained models like the Segment Anything Model \citep{kirillov2023segment}. Motivated by these advancements, our method echoes the unified model approach of \citep{graham2019hover}, but with specific enhancements tailored for our research objectives.

Inter-image variability inherent to histology data is another major difficulty in nuclear recognition. H\&E stained specimens sourced from different tissue types, fixation methods, or even hospitals can exhibit dramatic visual differences. Many publicly available datasets such as PanNuke \citep{gamper2019pannuke,gamper2020pannuke} and CoNSeP \citep{graham2019hover} comprise H\&E patches with relatively homogeneous styles since they draw data from only one or two sites. Models trained exclusively on such narrow datasets tend to overfit, rendering their ability to generalize to more diverse input lacking. Therefore, we opt to train our model on the CoNIC data \citep{graham2021conic}, the largest pathology dataset to date spanning diverse sources from 5 different institutions. By learning from such heterogeneous data, we aim to develop a segmentation and classification model with broader applicability across visual domains.

\subsection{Cellular Composition Prediction}
{Cellular composition prediction} in histology slides plays a crucial role in computational pathology, enabling analyses for various cancer types. For instance, accurately quantifying mitotic figures provides a critical parameter for tumor grade in breast cancer \citep{elston1991pathological,irshad2013methods}. Additionally, the percentage of tumor nuclei within the tumorbed area indicates the effect of neoadjuvant therapy in various cancers\citep{bersanelli2020tumour}. Computational composition analysis also sheds light on the tumor microenvironment (TME) \citep{galli2020relevance}, revealing the balance of malignant, immune, stromal, and healthy cell populations {coexisting} within the tumor ecology. 

The manual process of counting cells slide-by-slide is laborious and prone to subjective errors and variability across pathologists. The advent of digital pathology has enabled the development of automated computational approaches for cellular composition estimation aimed at supplementing pathologists' analyses. 

We can categorize existing research into two approaches: direct regression methods that output overall cell type counts without explicitly localizing individual nuclear instances \citep{khan2016deep,xue2016cell,dawood2021albrt}, and methods that first detect and segment all nuclei before tallying counts per category to derive composition \citep{xing2013automatic,cirecsan2013mitosis}. Since our goal is to develop an integrated model capable of both nuclear recognition and downstream composition prediction, we opt for the detection-based approach of first identifying nuclei via segmentation and classification, after which we can readily infer overall composition by counting nuclei in each predicted category.

\subsection{Ensemble Modeling}
Model ensembling has emerged as an effective technique for improving the performance of deep learning systems. The approach involves combining multiple models to create an ensemble system that {leverages} the complementary strengths of the individual models \citep{kotu2018data}. A key advantage of ensembling is reducing the generalization error and variability of predictions {\citep{kotu2018data}}. The base models comprising the ensemble are often trained separately, using distinct architectures, algorithms, or training data splits \citep{kotu2015data}. Though containing multiple components, the ensemble model essentially functions as a single unified model with lower aggregate error \citep{kotu2018data}.

Model ensembling has been successfully {applied} in many medical imaging applications. For example, \citep{ciresan2012deep} averaged the predictions of several networks to reduce variance in segmenting neuronal structures, winning the 2012 ISBI EM Segmentation Challenge. \citep{li2018fully} also utilized an ensemble of fully convolutional networks to detect white matter hyperintensities, achieving state-of-the-art performance in the 2017 MICCAI WMH Segmentation Challenge. Such results demonstrate the potential of ensemble techniques to boost performance. 

Common ensembling approaches include max voting, which selects the majority vote \citep{polikar2006ensemble,buhlmann2012bagging}, and averaging, which computes the mean predicted probabilities \citep{buhlmann2012bagging}. We implement an averaging ensemble, as it is widely adopted. Through designing ensemble algorithms tailored for our application, we aim to improve overall performance by synergistically combining multiple specialized base models.

\section{Methods}
Our proposed framework incorporates three vital components: 1) the base models for nucleus detection, 2) an ensemble system to combine base model outputs, and 3) post-processing techniques to refine ensemble predictions. At last, we describe the evaluation metrics and experimental setup.

\subsection{Base Models}
Our approach is built upon diverse base models, each employing a different encoder architecture as the backbone. Inspired by the HoVer-Net framework \citep{graham2019hover}, {which has achieved state-of-the-art performance on joint nuclear segmentation and classification tasks}, our base models share a common architecture, as illustrated in Fig. \ref{fig:base_model}. Each base model outputs three maps: 1) a nuclear pixel (NP) map predicting nuclei segmentation, 2) a horizontal/vertical distance (HoVer) map encoding the coordinates of nuclear centers, and 3) a nucleus classification (NC) map categorizing each nucleus. 

The NP map indicates whether each pixel belongs to a nuclear foreground region or the background. The HoVer map provides normalized horizontal and vertical distances to the centers of mass for clustered nuclei. By combining these two outputs, the model can separate touching and overlapping instances using a watershed transform during post-processing. Finally, the NC map classifies each segmented nucleus by performing pixel-wise voting within the predicted region. The category receiving the most votes defines the final call for that nucleus.

In our design, all three output maps share parameters within a common decoder pathway. By training these complementary maps jointly, we provide regularization for the decoder to learn robust features. We employ a single convolutional layer as the head for each map, minimizing added parameters. Our experiments validate that this simplified yet unified architecture achieves strong performance on both segmentation and classification. Moreover, by reducing computations, it allows us to use large batch sizes that accelerate training. 

We utilize two state-of-the-art models: SEResNeXt50 and SEResNeXt101 \citep{hu2019squeezeandexcitation}, as encoder backbones. We train a specialized base model for each encoder variant. Next, we detail our ensemble approach to synergistically combine the strengths of these diverse base models.

\subsubsection{{Model Regularization}}
To further enhance model generalization, we employ regularization strategies. As {illustrated} in Fig.~\ref{fig:base_model}A, we incorporate dropout layers periodically within the decoder after each upsampling layer. Dropout randomly omits or 'drops out' a subset of units during training, preventing the model from relying too heavily on particular features or cues that may vary across domains. This encourages the model to synthesize more holistic representations that do not overfit to nuances of the training distribution. Futhermore, our experiments demonstrate that the inclusion of dropout improves model performance on unseen test data.

\subsubsection{Loss Function}
Following HoVer-Net~\citep{graham2019hover}, we employ a composite multi-task loss to optimize the model:
\begin{equation}
   L = w_{NP}L_{NP} + w_{HoVer}L_{HoVer} + w_{NC}L_{NC},
\end{equation}
where $w_{NP}$, $w_{HoVer}$, and $w_{NC}$ are the weights for the different loss sets, and $L_{NP}$, $L_{HoVer}$, and $L_{NC}$ represent the loss terms for the nuclear pixel (NP) segmentation map, horizontal/vertical (HoVer) distance map, and nucleus classification (NC) map outputs respectively.

For the nuclear pixel (NP) branch, we apply a weighted combination of binary cross-entropy and dice losses. The binary cross-entropy loss provides pixel-level supervision for accurate nuclei vs. background classification. Meanwhile, the dice loss helps counter class imbalance between the typically smaller nuclear regions and the larger background area. Similarly, the nuclear classification (NC) branch is trained via analogous multi-task versions of cross-entropy and dice losses. Finally, for the horizontal/vertical distance (HoVer) branch, we utilize mean squared error and mean squared gradient error losses. These losses provide tailored supervision to precisely predict both the distance maps to nuclear centers themselves, as well as the distance gradient maps pointing to the centers. We refer the readers to the HoVer-Net~\citep{graham2019hover} paper for more details of the loss function design.

\subsection{Model Ensemble System}
\label{subsec:model_ensemble}
To harness the diversity of our trained base models, we propose a two-level model ensemble system {that} aggregates their predictions. The system consists of two levels: intra-model ensembling and inter-model ensembling. Intra-model ensembing combines predictions from differently augmented inputs with the same based model, while inter-model ensembling combines predictions from different base models. This approach is illustrated in Fig.~\ref{fig:base_model}B and Fig.~\ref{fig:base_model}C.

\subsubsection{Intra-Model Ensembling}
Intra-model ensembling, as illustated in Fig.~\ref{fig:base_model}B, combines the predictions from multiple augmented versions of the same base model. We generate augmented inputs by horizontally and vertically flipping the original input patch and feed these flipped variants through the base model to produce multiple output maps for the same image. We flip the nuclear pixel (NP) and nuclear classification (NC) maps back to their original orientation and average them to smooth predictions. For the horizontal/vertical (HoVer) distance maps, we additionally inverse the sign of values based on flip direction before averaging. This intra-model ensembling reduces errors and noise by aggregating consistent predictions.

\subsubsection{Inter-Model Ensembling}
Inter-model ensembling, as illustated in Fig.~\ref{fig:base_model}C, combines the predictions from multiple base models with distinct encoder backbones. Each base model undergoes intra-model ensembling to produce averaged output maps, which are then averaged across base models to generate the final prediction. By combining the complementary strengths of each base model, inter-model ensembling creates a more robust ensemble model that leverages the diversity of the individual models.

\subsection{Post-processing}
We employ post-processing techniques to get the final nuclear instance segmentations and classifications. The process involves two primary steps: watershed segmentation and pixel-wise voting. The HoVer and NP maps serve as inputs to the watershed algorithm, where the HoVer map defines an energy landscape with nuclear centers as catchment basins, and the NP map removes background areas by zeroing their HoVer distances. Following segmentation, each nucleus is classified via pixel-wise voting within the nuclear instance region, assigning the most frequent class. By leveraging all three output maps in concert, the model is able to perform end-to-end detection, segmentation, and classification.

The workflow is illustrated in Fig.~\ref{fig:base_model}D, where the HoVer and the NP maps are utilized for instance segmentation, and the NC map is used for classification. The resulting segmented and classified instances are then used to count the number of nuclei of each category, enabling the prediction of nuclear composition within the central $224\times224$ region for image patches as shown in Fig.~\ref{fig:base_model}E.

\subsection{Datasets} 
The quality and diversity of training data are crucial for developing robust computational pathology models. Histopathology images exhibit high variability due to factors such as tissue type, fixation method, and source institution. To mitigate dataset bias and train more generalizable models, we leverage multi-site datasets with heterogeneous visual styles for pre-training. We opt to train on the large-scale Lizard repository, which contains annotated pathology images aggregated from 6 unique domains, with iterative pathologist input to refine labels. We also validate our method on two other datasets: PanNuke and MoNuSAC. Detailed information on these datasets is illustrated in Table~\ref{tab:table_datasets}.

\begin{table}[t!]
\centering
\caption{Detailed information of the datasets. We provide a detailed overview of the three public datasets, including the number of patches, nuclei, and classes, to illustrate the composition of the datasets.}
\resizebox{0.48\textwidth}{!}{
 \begin{tabular}{c|ccc}
 \toprule
        &No. Patches &No. Nuclei &Classes \\
\hline
\makecell{MoNuSAC\\ \citep{verma2021monusac2020}}
 &310 &$\approx$46,000 &\makecell{epithelial, lymphocyte, \\neutrophil, macrophage}\\
  \hline
  \makecell{PanNuke\\ \citep{gamper2019pannuke,gamper2020pannuke}}
 &7901 &$\approx$200,000 &\makecell{neoplastic, inflammatory, \\
                                connective, dead, epithelial}\\
\hline
\makecell{Lizard\\ \citep{graham2021lizard}}
    &4981 &$\approx$500,000 &\makecell{epithelial, lymphocyte, \\
                            plasma, eosinophil, \\
                            neutrophil, connective tissue}\\
\hline
 \end{tabular}
 }
 \label{tab:table_datasets}
\end{table}

\subsubsection{MoNuSAC}
The MoNuSAC dataset \citep{verma2021monusac2020} contains over 46,000 nuclei across 37 hospitals, 71 patients, 4 organs, and 4 {nuclear} classes - epithelial, lymphocyte, neutrophil, and macrophage. Sourced from TCGA, the whole slide images are sampled at 40x magnification. Annotations were performed by engineering graduate students and quality was checked by domain experts. Considering that the MoNuSAC dataset \citep{verma2021monusac2020} does not provide additional data splits in its original publication, we chose not to perform cross-validation on it. Instead, we present the mean and standard deviation of 10 experimental runs, ensuring a consistent approach that facilitates direct comparison of performance metrics across all datasets.

\subsubsection{PanNuke}
The PanNuke \citep{gamper2019pannuke,gamper2020pannuke} dataset contains multi-organ data across 19 tissue types with nearly 200,000 annotated nuclei in 5 clinically relevant classes - neoplastic, inflammatory, connective, dead, and epithelial. It consists of 481 visual fields sampled at 40x magnification where nuclei are semi-automatically segmented and verified by experts. To facilitate robust model evaluation, the dataset is pre-split into three folds. We adhere to the original split of the PanNuke dataset to perform cross-validation, as detailed in Table~\ref{tab:table_data_fold}. Note that all other results for PanNuke presented in this paper are based on Fold 0, with mean and standard deviation calculated from 10 independent experimental runs, maintaining uniformity in our dataset evaluations.

\subsubsection{Lizard}
The Lizard dataset \citep{graham2021lizard} contains colon tissue samples sourced from 5 distinct repositories - Glas \citep{sirinukunwattana2017gland}, CRAG \citep{graham2019mild}, CoNSep \citep{graham2019hover}, DigestPath, PanNuke \citep{gamper2019pannuke,gamper2020pannuke}. It is used by the CONIC 2022 challenge, with the composition detailed in an excel file provided by the challenge organizers~\citep{graham2021conic}\footnote{https://conic-challenge.grand-challenge.org/}. To maintain the integrity of the original sources and to align with the cross-validation approach, we split the Lizard dataset into five folds corresponding to these sources to perform cross-validation, as shown in Table~\ref{tab:table_data_fold}. Note that all other results for Lizard presented in this paper are performed on Fold 0, with mean and standard variations calculated from 10 times of experiments, as for MoNuSAC and PanNuke.

\begin{figure}[t!]
\centering
\begin{subfigure}{0.13\textwidth}
  \centering
  \includegraphics[width=\textwidth]{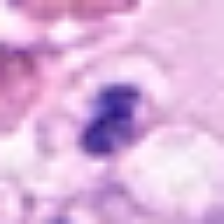}
  \caption{Neutrophil}
  \label{fig:dataset_neutrophil}
\end{subfigure}
\hspace{3pt}
\begin{subfigure}{0.13\textwidth}
  \centering
  \includegraphics[width=\textwidth]{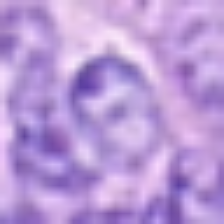}
  \caption{Epithelial}
  \label{fig:dataset_epithelial}
\end{subfigure}
\hspace{3pt}
\begin{subfigure}{0.13\textwidth}
  \centering
  \includegraphics[width=\textwidth]{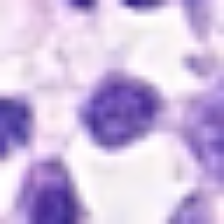}
  \caption{Lymphocyte}
  \label{fig:dataset_lymphocyte} 
\end{subfigure} 
\\ 
\vspace{10pt}
\begin{subfigure}{0.13\textwidth}
  \centering
  \includegraphics[width=\textwidth]{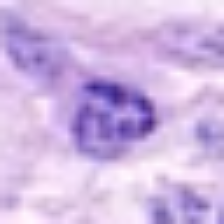}
  \caption{Plasma}
  \label{fig:dataset_plasma}
\end{subfigure}
\hspace{3pt}
\begin{subfigure}{0.13\textwidth}
  \centering
  \includegraphics[width=\textwidth]{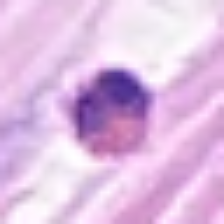}
  \caption{Eosinophil}
  \label{fig:dataset_eosinophil} 
\end{subfigure} 
\hspace{3pt}
\begin{subfigure}{0.13\textwidth}
  \centering
  \includegraphics[width=\textwidth]{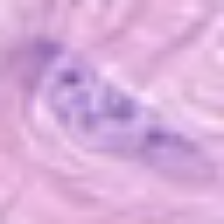}
  \caption{Connective}
  \label{fig:dataset_connective}
\end{subfigure}
\caption{Example nuclei from the 6 categories in the Lizard dataset, illustrating the challenges in segmentation and classification. {Note the high degree of similarity among certain classes (e.g., Fig.~\ref{fig:Lizard_dataset}(b) Epithelial,  Fig.~\ref{fig:Lizard_dataset}(c) Lymphocyte, and Fig.~\ref{fig:Lizard_dataset}(d) Plasma), which hinders classification.}}
\label{fig:Lizard_dataset}
\end{figure}

The dataset is cropped into 4981 256$\times$256 patches containing around 500,000 annotated nuclei. Figure~\ref{fig:Lizard_dataset} illustrates example patches from each nucleus category to provide intuition. While images of colon tissues from PanNuke are included in Lizard, we evaluated these two datasets independently without any overlap between training and test sets to avoid information leakage. Regions of interest are extracted from the whole slide images at 20x magnification and nuclei are annotated into 6 categories - epithelial, lymphocyte, plasma, eosinophil, neutrophil, and connective tissue.

\begin{figure}[t!]
\centering
\includegraphics[width=0.5\textwidth]{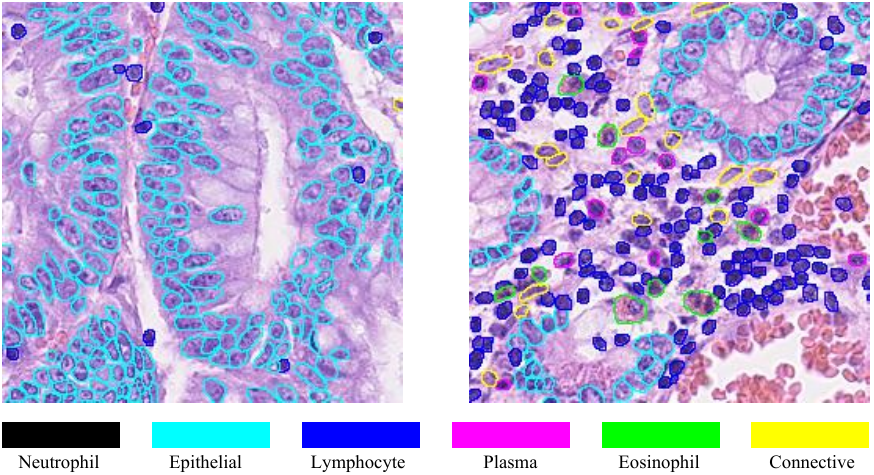}
\caption{{Example images from the Lizard dataset. Note the high degree of nuclei overlapping, which poses a significant challenge for instance segmentation methods.}}
\label{fig:overlapping}
\end{figure}

\begin{table*}[t!]
\centering
\caption{Performance comparison of the proposed framework against other methods on the nuclear recognition and composition prediction tasks over multiple datasets. The proposed framework consistently outperforms the other methods by a substantial margin in most cases. Even in the rare instances where our approach does not surpass all others, it remains comparable to the best models. This demonstrates the effectiveness of our method for nuclear recognition and quantification. Further analysis of the results is provided in Section \ref{subsec:comparison_with_others}.}
\resizebox{\textwidth}{!}{
 \begin{tabular}{c|ccc|ccc}
 \toprule
&\multicolumn{3}{c|}{$mPQ+$}
&\multicolumn{3}{c}{$R^2$}\\
\hline
        &Lizard &PanNuke &MoNuSAC &Lizard &PanNuke &MoNuSAC  \\
\hline
Mask R-CNN~\citep{he2017mask}
    &0.4281$\pm$0.0110 &0.4024$\pm$0.0246 &0.5072$\pm$0.0069 
    &0.4459$\pm$0.0233 &0.6771$\pm$0.1016 &0.6105$\pm$0.0187\\

Cascaded R-CNN~\citep{Cai_2019}
&0.4493$\pm$0.0077 &0.4420$\pm$0.0125 &0.5278$\pm$0.0027
&0.4907$\pm$0.0214 &\textbf{0.7640$\pm$0.0201} &0.6484$\pm$0.0272\\

QueryInst~\citep{Fang_2021_ICCV}
&0.3549$\pm$0.0017 &0.4405$\pm$0.0031 &0.4938$\pm$0.0149 
&0.4552$\pm$0.0049 &0.7134$\pm$0.0147 &0.5813$\pm$0.2020 \\

StarDist~\citep{weigert2020}
&0.4194$\pm$0.0128 &0.3855$\pm$0.0108 &0.4097$\pm$0.0068 
&0.4795$\pm$0.0470 &0.5662$\pm$0.0236 &0.6774$\pm$0.1475 \\

HoVer-Net~\citep{graham2019hover} 
&0.4893$\pm$0.0217 &0.3736$\pm$0.0111 &0.4501$\pm$0.0154 
&0.7081$\pm$0.1148 &0.5913$\pm$0.0237 &0.5933$\pm$0.1419 \\

Ours &\textbf{0.5599$\pm$0.0086} &\textbf{0.4738$\pm$0.0061} &\textbf{0.5567$\pm$0.0125} 
     &\textbf{0.8437$\pm$0.0454} &0.7127$\pm$0.0553 &\textbf{0.7968$\pm$0.0470} \\
 \hline
 \end{tabular}
 }
 \label{tab:table_comparison}
\end{table*}

\section{Experiments}
\label{sec:experiments}
We conduct extensive experiments to evaluate the performance of our proposed framework. This section describes the evaluation metrics and experimental setup.

\subsection{Evaluation Metrics}
\label{subsec:evaluation_metrics}
We evaluate nuclei detection, segmentation, and composition estimation using two key metrics{:} $mPQ+$ for nuclear instance recognition and classification and $R^2$ for nuclear composition regression.  

The $mPQ+$ metric evaluates combined nuclear segmentation and classification performance, adapted from the Panoptic Quality (PQ) metric \citep{kirillov2019panoptic} introduced in \citep{graham2019hover} for computational pathology tasks. Specifically, for each nuclear type $t$, $PQ_{t}$ is defined as:

\begin{equation}
    PQ_{t} = \frac{|{TP}_{t}|}{|{TP}_{t}|+\frac{1}{2}|{FP}_{t}|+\frac{1}{2}|{FN}_{t}|} \times \frac{\Sigma_{(x_t,y_t)\in {TP}_{t}}IoU(x_t,y_t)}{|{TP}_{t}|}
\end{equation}

Here, $x_t$ refers to a ground truth nucleus of type $t$, $y_t$ refers to a predicted nucleus, and $IoU(x_t,y_t) > 0.5$ for matched nuclei pairs $(x_t,y_t)\in {TP}_{t}$. Unmatched ground truth and predicted nuclei are counted as false negatives ($FN_{t}$) and false positives ($FP_{t}$) respectively. 

{The $mPQ+$ metric evaluates segmentation and classification performance by averaging the $PQ_{t}$ score across all nuclear types $t$. Critically, $PQ_{t}$ is calculated over the full dataset rather than individual images, addressing potential issues that may arise from zero denominators when classes are absent from some images. For full details on the $mPQ+$ formulation and its advantages over standard $mPQ$, we direct readers to the work presenting this evaluation framework in the context of the CoNIC Challenge~\citep{graham2021conic}.}

For quantitative composition estimation, we report the $R^{2}$ coefficient of determination. For each nuclear type $t$, $R^{2}_{t}$ is:

\begin{equation}
    R^2_{t} = 1 - \frac{\Sigma_{i=1}^{n}(y_{i,t} - {\hat{y}_{i,t}})}{\Sigma_{i=1}^{n}(y_{i,t} - \Bar{y}_{t})^{2}}
\end{equation}

Here, $y_{i,t}$ is the ground truth count for nuclei of type $t$ in image patch $i$, $\hat{y}_{i,t}$ is the predicted count, and  $\Bar{y}_{t}$ is the mean $y_{i,t}$ across patches. The overall multi-class $R^{2}$ score averages the per-type $R^2_t$ metrics.

As illustrated in Fig.~\ref{fig:base_model}E, $mPQ+$ is computed over the full image patch ($256\times256$), while $R^{2}$ uses only the central $224\times224$ region to avoid edge effects. 

\begin{table}[t!]
\centering
\caption{Comparison of the top 10 teams on the CoNIC 2022 challenge leaderboard for the two evaluation metrics. Our proposed approach achieved 3rd place for multi-class panoptic quality ($mPQ+$) and 1st place for composition regression ($R^2$). The blind test set indicates the superiority of our method on unseen custom data.}
\resizebox{0.48\textwidth}{!}{
 \begin{tabular}{cc|cc}
 \toprule
 Team &$mPQ+$ &Team &$R^2$\\
 \hline
EPFL \textbar\, StarDist     &0.50132
&\textbf{Pathology AI}       &\textbf{0.76413} \\

MDC Berlin \textbar\, IFP Bern &0.47616 
&AI\_medical                 &0.76250 \\

\textbf{Pathology AI}        &\textbf{0.46310}
&EPFL \textbar\, StarDist    &0.75498 \\

LSL000UD                     &0.46278
&CIA Group                   &0.71902 \\

AI\_medical                  &0.45759
&Softsensor\_Group           &0.71589 \\

Arontier                     &0.45707
&LSL000UD                    &0.70325 \\

CIA Group                    &0.45092
&Arontier                    &0.69145 \\

MAIIA                        &0.43674
&MBZUAI\_CoNiC               &0.67440 \\

ciscNet                            &0.42947
&MDC Berlin \textbar\, IFP Bern    &0.66432 \\

MBZUAI\_CoNiC                 &0.42078
&Denominator                 &0.65498 \\
\hline
 \end{tabular}
 }
 \label{tab:table_challenge}
\end{table}

\subsection{Implementation Details}
\label{subsec:implementation_details}

% Encoders
In this work, we employ two powerful encoder backbones, {SEResNeXt50} and {SEResNeXt101} \citep{hu2019squeezeandexcitation}, to extract robust features in our multi-task segmentation framework. The SEResNeXt models are built upon the ResNeXt architecture and incorporate Squeeze-and-Excitation blocks to improve feature representation. During inference, {we ensemble the predictions of both models to obtain the final segmentation result. By combining diverse models, we can leverage their strengths and} improve robustness and accuracy.

Our training is performed on a high-performance workstation with 4 NVIDIA 3090 GPUs using distributed parallel training. This allows us to train with a large batch size of 16 for faster convergence. Models are trained for 50 epochs each, using the Adam optimizer \citep{kingma2014adam} with an initial learning rate of 3e-4. The learning rate is decayed by a factor of 0.1 every 10 epochs after the first 30 epochs to improve convergence at the end of training. To prevent overfitting, we employ an early stopping strategy based on the validation loss in the challenge. Training stops once the loss stabilizes and stops improving, and we use the model from several epochs prior.

To enhance robustness and prevent overfitting, we apply extensive data augmentation techniques to our training set. These include random horizontal/vertical flips, random 90-degree rotations, random transposing, color jittering of brightness, contrast, saturation and hue, Gaussian blurring, median blurring, and motion blurring of the images. We only apply one blurring operation each time to avoid corrupting the image too much. 

{We use a weighted combination of loss components from each branch. Through a grid search, we found that} $w_{NP}$, $w_{HoVer}$, and $w_{NC}$=1 work best. The NP branch uses $w_{CE}$=2 and $w_{Dice}$=2 for its cross entropy and Dice loss components. The hover branch uses $w_{mCE}$=3 and $w_{mDice}$=1 to weight its modified cross entropy and Dice losses. The keypoint loss uses $w_{mse}$=2 and $w_{msge}$=2 for the MSE and MSGE losses. Carefully balancing the loss components prevents one task from dominating the others during training.

\begin{table*}[t!]
\centering
\caption{Performance comparison between our proposed framework and the original HoVer-Net \citep{graham2019hover} on the Lizard dataset \citep{graham2021lizard}. Our approach consistently and significantly outperforms HoVer-Net on these categories on both evaluation metrics, demonstrating the advantages of our method.}
\resizebox{\textwidth}{!}{
 \begin{tabular}{ccccccc|cccccc}
 \toprule
 &\multicolumn{6}{c|}{$mPQ+$}
&\multicolumn{6}{c}{$R^2$}\\
\hline
&Neutrophil &Epithelial &Lymphocyte &Plasma &Eosinophil &Connective
&Neutrophil &Epithelial &Lymphocyte &Plasma &Eosinophil &Connective\\
 \hline
HoVer-Net 
&0.2987$\pm$0.0399 &0.6216$\pm$0.0213 &0.7108$\pm$0.0196 
&0.4611$\pm$0.0177 &0.3085$\pm$0.0524 &0.5353$\pm$0.0084 
&0.8980$\pm$0.0436 &0.9470$\pm$0.0244 &0.5768$\pm$0.2246 
&0.7331$\pm$0.1057 &0.6553$\pm$0.1481 &0.4387$\pm$0.3953 \\
Ours      
&0.3941$\pm$0.0227 &0.6684$\pm$0.0028 &0.7503$\pm$0.0047 
&0.5180$\pm$0.0114 &0.3980$\pm$0.0218 &0.6305$\pm$0.0084 
&0.9586$\pm$0.0126 &0.9847$\pm$0.0048 &0.7531$\pm$0.0495 
&0.7147$\pm$0.2430 &0.7881$\pm$0.0450 &0.8628$\pm$0.0393 \\
\hline
 \end{tabular}
 }
 \label{tab:table_per_category}
\end{table*}

\section{Results}
\label{sec:results}
In this section, we present experimental results comparing our proposed multi-task segmentation framework against current state-of-the-art methods on two key tasks: 1) {Nuclear instance segmentation and classification}, and 2) Nuclear composition regression, as defined in the CoNIC 2022 Challenge \citep{graham2021conic}. We also use Dice to calculate the model's performance on segmentation. Ablation studies are conducted to investigate the contribution of each component of our framework.

\subsection{Comparison With Other State-of-the-Art Methods}
\label{subsec:comparison_with_others}
We compare our approach to several state-of-the-art methods on nuclear instance segmentation and classification, including Mask R-CNN \citep{he2017mask}, Cascaded R-CNN \citep{Cai_2019}, and QueryInst \citep{Fang_2021_ICCV}. We also compare against HoVer-Net \citep{graham2019hover} and star-dist~\citep{schmidt2018,weigert2020,weigert2022}, two specifically designed methods for nuclear instance segmentation and classification. We show the quantitative results in Table \ref{tab:table_comparison}, reporting standard evaluation metrics such as segmentation and classification ($mPQ+$), {as well as} nuclear composition regression ($R^2$ score). Our framework demonstrates superior performance across the majority of datasets and evaluation metrics, showcasing its robustness and versatility. In instances where our model does not lead in a particular metric, it remains highly competitive, exhibiting performance on par with the state-of-the-art methods. A case in point is the $R^2$ score for the PanNuke dataset, where our model, despite not achieving the foremost position, closely matches the performance of the leading cascaded R-CNN approach.

We have also conducted a comparative analysis of our method with the composite work CellViT~\citep{horst2024cellvit}, utilizing the metrics employed in its original publication: $mPQ$ and $bPQ$, which are variants of our enhanced metric, $mPQ+$. For a comprehensive understanding of these metrics, we direct the readers to the CellViT paper for detailed explanations.

\begin{table}[t!]
\centering
\caption{Performance comparison between our method and CellViT. The results indicate that our method, which does not leverage large-scale pretrained models, demonstrates performance that is highly comparable to CellViT.}
\resizebox{0.45\textwidth}{!}{
  \begin{tabular}{c|cc|c}
 \toprule
&$CellViT_{256}$ &CellViT-SAM-H &Ours \\
 \hline
$mPQ$  &0.4846$\pm$0.0503 &0.4980$\pm$0.0413 &0.5060$\pm$0.0022\\
$bPQ$  &0.6696$\pm$0.0340 &0.6793$\pm$0.0318 &0.6353$\pm$0.0065\\
\hline 
 \end{tabular}
 }
 \label{tab:table_cellvit}
\end{table}

\begin{figure*}[t!]
\centering
\includegraphics[width=\textwidth]{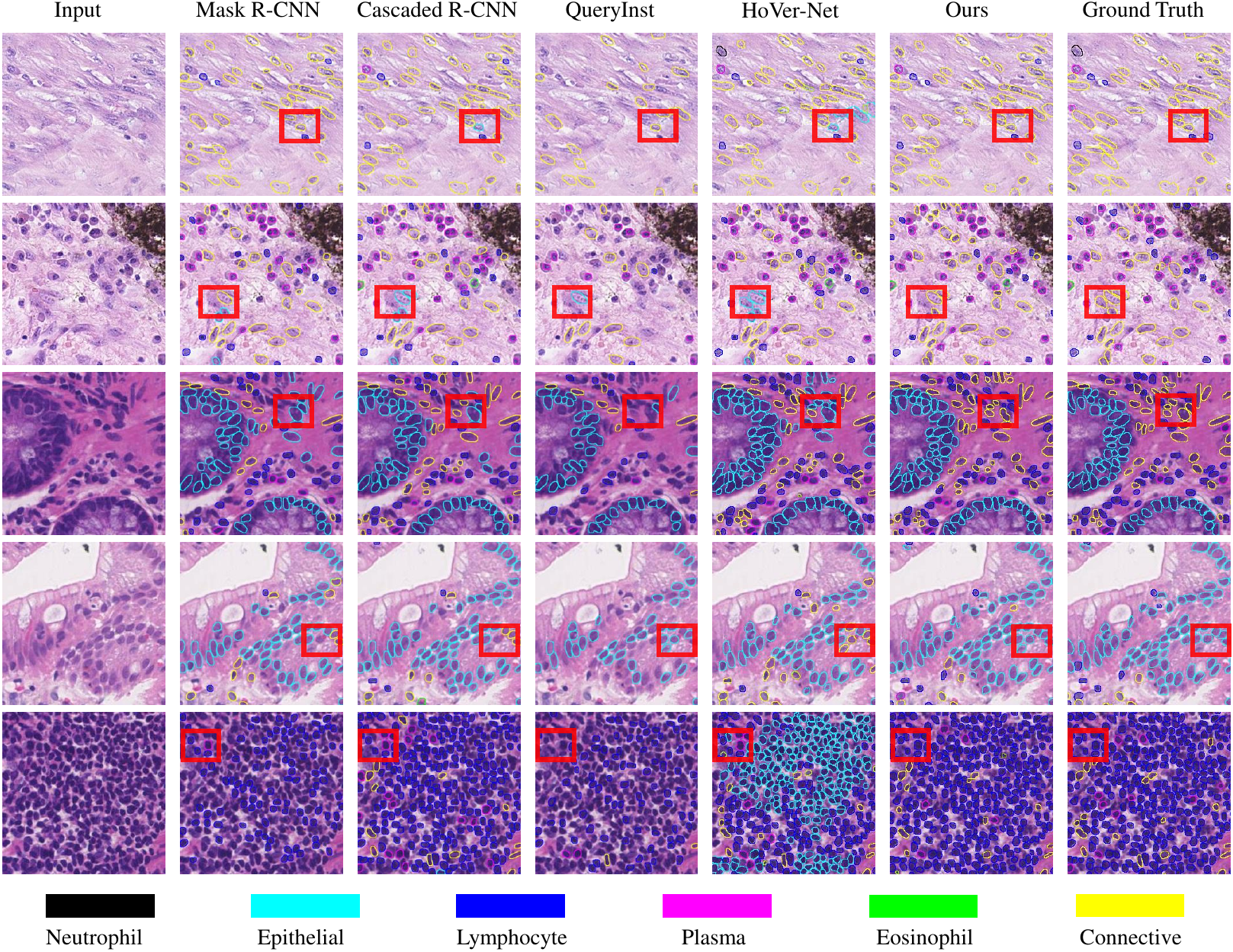}
\caption{Qualitative results comparing the proposed method against other state-of-the-art approaches on example images from the Lizard dataset \citep{graham2021lizard}. The overlays visualize the output instance segmentation results, with different colors indicating the predicted nuclear categories. Examples of nuclei from each category are provided in Fig. \ref{fig:Lizard_dataset}. The proposed method produces more accurate predictions than other state-of-the-art techniques, as observed by examining the overlays. Our approach correctly identifies more instances, with fewer false positives and false negatives. Additionally, the nuclear category predictions match the ground truth more closely compared to other methods. This qualitative analysis highlights the improved performance of the proposed framework on this challenging nuclear recognition task. }
\label{fig:results_vis}
\end{figure*}

{We showcase visual results on the Lizard dataset in Fig. \ref{fig:results_vis}, highlighting cleaner and more accurate segmentations and classifications compared to other methods. Our multi-task approach appears to learn a robust feature representation that captures distinguishing characteristics between classes. Furthermore, the ensemble ability of our model allows it to incorporate information that may be overlooked in single models. Additionally, we break down results by each cell category in the Lizard dataset in Table \ref{tab:table_per_category}, showing top scores in every category for both $mPQ+$ and $R^2$ metrics compared to the HoVer-Net baseline. These consistent gains demonstrate the broad improvements from our proposed approach and validate the utility of our architecture modifications.}

In summary, both quantitative metrics and qualitative results confirm the effectiveness of our multi-task framework relative to state-of-the-art methods for nuclear instance segmentation, classification, and composition regression. The joint training provides complementary information that boosts performance across related tasks.

\subsection{Comparison to Top Teams in CoNIC 2022 Challenge}
We participated in the 2022 Colon Nuclei Identification and Counting Challenge (CoNIC 2022 Challenge) organized by \citep{graham2021conic}, a competition aimed at advancing research on automatic nuclear instance characterization for computational pathology applications. {The challenge involved segmenting}, classifying, and counting 6 nuclear types in provided multi-organ datasets. Blinded test data, including samples from new domains not seen during training, posed a significant challenge. To generalize well to diverse data, high-performing methods were needed.

Submissions were evaluated based on mean panoptic quality ($mPQ+$) for segmentation and classification, and $R^2$ score for nuclear composition regression. Table \ref{tab:table_challenge} {shows} the final scores of the top 10 ranked teams on the test set. The top methods achieved very close scores, indicating highly competitive solutions were developed by leading research groups. Our proposed multi-task framework achieved  \textbf{1st place} for the nuclear composition $R^2$ metric with a score of 0.76413, and 3rd place based on the $mPQ+$ metric with a score of 0.46310.

This demonstrates our method's ability to generalize successfully to new datasets while maintaining top-tier performance relative to state-of-the-art techniques. The joint multi-task learning provides our model with robust representations that transfer well to unseen domains. Our strong ranking in this competitive challenge confirms the effectiveness of the proposed approach for real-world biomedical image analysis tasks requiring generalization.

\begin{table*}[t!]
\centering
\caption{Evaluation of base models utilizing different encoder backbones. As shown, larger encoder models tend to achieve better or at least comparable performance to smaller models. {Consistent with the findings in nnUNet-Revisited \citep{isensee2024nnu}, we also observe that Transformer-based architectures do not surpass the performance of CNNs in our task.} Thus, we employ SEResNeXt50 \citep{hu2018squeeze} and SEResNeXt101 \citep{hu2018squeeze} encoders in our framework.} 
\resizebox{0.9\textwidth}{!}{
  \begin{tabular}{cccc|ccc}
 \toprule
&\multicolumn{3}{c|}{$mPQ+$}
&\multicolumn{3}{c}{$R^2$}\\
\hline
             &Lizard &PanNuke &MoNuSAC &Lizard &PanNuke &MoNuSAC\\
 \hline
ResNet50     
&0.4869$\pm$0.0234 &0.3437$\pm$0.0372 &0.4569$\pm$0.0197 
&0.6921$\pm$0.1254 &0.5560$\pm$0.0613 &0.6356$\pm$0.1199 \\
ResNet101    
&0.4980$\pm$0.0279 &0.3902$\pm$0.0194 &0.4686$\pm$0.0233 
&0.7313$\pm$0.1087 &0.6378$\pm$0.0516 &0.5487$\pm$0.1744 \\
{SEResNeXt50}
&0.5087$\pm$0.0327 &0.4322$\pm$0.0216 &0.4712$\pm$0.0156 
&0.7529$\pm$0.1143 &0.6658$\pm$0.0622 &\textbf{0.6356$\pm$0.0906}\\
{SEResNeXt101} 
&\textbf{0.5286$\pm$0.0373} &\textbf{0.4489$\pm$0.019}6 &\textbf{0.4815$\pm$0.014}2 
&\textbf{0.7700$\pm$0.1667} &\textbf{0.6942$\pm$0.0479} &0.6288$\pm$0.1013 \\
Swin-Transformer 
&0.4544$\pm$0.0683 &0.3030$\pm$0.0203 &0.4480$\pm$0.0062 
&0.6547$\pm$0.0642 &0.6176$\pm$0.0957 &0.5839$\pm$0.0631 \\
\hline
 \end{tabular}
 }
 \label{tab:table_encoder_structure}
\end{table*}

\subsection{Ablation Study}
\label{subsec:ablation_study}
{To investigate the importance of individual components in our proposed framework, we conduct an ablation study. We analyze the contributions of four key components: encoder backbones, data augmentation strategies, class imbalance handling methods, and model ensemble techniques. Understanding the individual contributions of these components is essential for further improving the model.}

\subsubsection{Encoder Backbones}
\label{subsubsec:ablation_encoder_backbones}

{The performance of the encoder is a crucial factor in determining the overall success of the model}. We experiment with various encoder architectures, as listed in Table \ref{tab:table_encoder_structure}, including ResNet50, ResNet101 \citep{he2016deep}, SEResNeXt50, SEResNeXt101 \citep{hu2018squeeze}, and swin-transformer~\citep{swinunet,liu2021Swin}. The SEResNeXt models incorporate ResNeXts with squeeze-and-excitation blocks to enhance representation learning, {while the swin-transformer leverages attention mechanisms to emphasize important features.}

{Our results indicate that} larger models, such as SEResNeXt101, tend to perform better due to their increased capacity to model complex patterns. However, this {trend} is not absolute, as {evident from the slight outperformance of ResNet101 on the $R^2$ metric}. {Furthermore}, very large models can overfit on the limited training data. {Notably, our findings align with those in nnUNet-Revisited~\citep{isensee2024nnu}, suggesting that Transformer-based architectures do not surpass the performance of CNNs in our task.} Based on these findings, we select SEResNeXt50 and {SEResNeXt101} as our encoder backbones, as they achieve top performance on most datasets.

\begin{table}[t!]
\centering
\caption{{Impact on the performance of different data augmentation techniques applied during training using the SEResNeXt50 \citep{hu2018squeeze} encoder. Models trained with most augmentations outperform no augmentation, except distortion which is unstable. Thus, distortion is excluded from our final training pipeline.}}
\resizebox{0.4\textwidth}{!}{
  \begin{tabular}{c|cc}
 \toprule
&$mPQ+$ &$R^2$ \\
 \hline
No Augmentation      &0.5300$\pm$0.0128 &0.7758$\pm$0.0318\\
Flip                 &0.5339$\pm$0.0126 &0.7739$\pm$0.0288\\
Color Jitter         &0.5382$\pm$0.0102 &0.7962$\pm$0.0295\\
Blur                 &0.5356$\pm$0.0091 &0.7915$\pm$0.0273\\
Distort              &0.5198$\pm$0.0084 &0.7628$\pm$0.0275\\
\hline
 \end{tabular}
 }
 \label{tab:table_augmentation}
\end{table}

\begin{table}[t!]
\centering
\caption{{Evaluation of dropout as a model regularization strategy. Dropout layers are included to prevent the model from developing an over-reliance on particular features, which could lead to increased sensitivity to variations in unseen test data.}}
\resizebox{0.4\textwidth}{!}{
  \begin{tabular}{c|ccc}
 \toprule
&$mPQ+$ &$R^2$\\
 \hline
Without dropout             &0.5504$\pm$0.0125 &0.8394$\pm$0.0305\\
With dropout             &0.5599$\pm$0.0086 &0.8437$\pm$0.0454\\
\hline
 \end{tabular}
 }
 \label{tab:table_regularization}
\end{table}

\begin{table}[t!]
\centering
\caption{{Evaluation of imbalanced class techniques. We explored various techniques to address class imbalance but found that they did not significantly improve performance. Therefore, we did not incorporate them into our method.}}
\resizebox{0.4\textwidth}{!}{
  \begin{tabular}{c|ccc}
 \toprule
&$mPQ+$ &$R^2$\\
 \hline
Original               &0.5599$\pm$0.0086 &0.8437$\pm$0.0454\\
Focal Loss             &0.5431$\pm$0.0112 &0.8159$\pm$0.0306\\
Weighted Cross-Entropy  &0.5520$\pm$0.0082 &0.8374$\pm$0.0255\\
\hline
 \end{tabular}
 }
 \label{tab:table_data_imbalance}
\end{table}

\begin{table}[t!]
\centering
\caption{Performance comparison across folds for Lizard and PanNuke datasets. The table details the performance of our model across the five folds of the Lizard dataset and the three folds of the PanNuke dataset. The Lizard dataset's composition, stemming from five distinct sources, justifies its division into separate folds for a comprehensive evaluation.  For the PanNuke dataset, we adhered to the original data split as described in its publication. We excluded MoNuSAC from this analysis due to its singular test dataset, aligning with the original split for consistency with other studies.}
\resizebox{0.5\textwidth}{!}{
  \begin{tabular}{c|c|ccc}
 \toprule
Dataset
& Split
&{$Dice$}
&{$mPQ+$}
&{$R^2$} \\
 \hline
\multirow{5}{*}{Lizard}
&Fold 0  &0.6557$\pm$0.0259 &0.5599$\pm$0.0086 &0.8437$\pm$0.0454\\
&Fold 1  &0.6270$\pm$0.0185 &0.5172$\pm$0.0096 &0.7211$\pm$0.1257\\
&Fold 2  &0.4779$\pm$0.0204 &0.4768$\pm$0.0114 &0.5982$\pm$0.0654\\
&Fold 3  &0.6362$\pm$0.0223 &0.5456$\pm$0.0096 &0.7998$\pm$0.0941\\
&Fold 4  &0.6483$\pm$0.0088 &0.5603$\pm$0.0076 &0.8366$\pm$0.0402\\
\hline
\multirow{3}{*}{PanNuke}
&Fold 0  &0.6355$\pm$0.0058 &0.4738$\pm$0.0061 &0.7127$\pm$0.0553
\\
&Fold 1  &0.6428$\pm$0.0102 &0.4875$\pm$0.0099 &0.7600$\pm$0.0658\\
&Fold 2  &0.6442$\pm$0.0155 &0.4892$\pm$0.0117 &0.8041$\pm$0.0292\\
\hline
 \end{tabular}
 }
 \label{tab:table_data_fold}
\end{table}
\subsection{Data Augmentation}
\label{subsec:ablation_data_augmentation}
{Data augmentation is a powerful tool for improving the generalization of our model}. We carefully select augmentations suitable for H\&E histology images to avoid distorting important color and texture cues. As shown in Table \ref{tab:table_augmentation}, most techniques such as {1) random flips with a probability of 0.5, 2) random rotation with a probability of 0.5, 3) color jittering with brightness jitter of 0.2, contrast jitter of 0.25, saturation jitter of 0.2, and hue jitter of 0.5, 4) Gaussian blur with a blur limit of 3, 5) Median blur with a blur limit of 3,} improve performance of our model by introducing more variation. However, {we also observe that distortion augmentation can sometimes decrease performance metrics}. This may due to over-augmentation. Thus, we exclude distortion and {focus on other effective augmentation techniques to optimize our model's performance}.

\begin{table*}[t!]
\centering
\caption{Evaluation of intra-model and inter-model ensemble techniques. {“Intra-model” employs solely intra-model ensembling, and “Inter-Model” utilizes only inter-model ensembling. “Ours” combines both intra-model and inter-model ensembling strategies. The data demonstrates that ensemble methodologies consistently enhance models’ performance.}}
\resizebox{0.8\textwidth}{!}{
  \begin{tabular}{cccc|ccc}
 \toprule
&\multicolumn{3}{c|}{$mPQ+$}
&\multicolumn{3}{c}{$R^2$}\\
\hline
             &Lizard &PanNuke &MoNuSAC &Lizard &PanNuke &MoNuSAC \\
 \hline
 {SEResNeXt50} 
&0.5087$\pm$0.0327 &0.4322$\pm$0.0216 &0.4712$\pm$0.0156 
&0.7529$\pm$0.1143 &0.6658$\pm$0.0622 &0.6356$\pm$0.0906 
\\
{SEResNeXt50 (Intra-model)}
&0.5488$\pm$0.0063 &0.4326$\pm$0.0218 &0.4916$\pm$0.0211 
&0.8455$\pm$0.0270 &0.6659$\pm$0.0630 &0.6988$\pm$0.0885 
\\
{SEResNeXt101}  
&0.5286$\pm$0.0373 &0.4489$\pm$0.0196 &0.4815$\pm$0.0142 
&0.7700$\pm$0.1667 &0.6942$\pm$0.0479 &0.6288$\pm$0.1013  
\\
{SEResNeXt101 (Intra-model)}  
&0.5482$\pm$0.0264 &0.4506$\pm$0.0185 &0.5011$\pm$0.0228 
&0.8009$\pm$0.1796 &0.6955$\pm$0.0481 &0.6904$\pm$0.1020
\\
{Inter-model}  
&{0.5352$\pm$0.0159} &{0.4686$\pm$0.0082} &{0.5101$\pm$0.0255} 
&{0.8185$\pm$0.0361} &{\textbf{0.7260$\pm$0.0528}} &{0.7134$\pm$0.1027}\\
Ours (ensemble all) 
&\textbf{0.5599$\pm$0.0086} &\textbf{0.4738$\pm$0.0061} &\textbf{0.5567$\pm$0.0125} 
&\textbf{0.8437$\pm$0.0454} &0.7127$\pm$0.0553 &\textbf{0.7968$\pm$0.0470} 
\\
\hline
 \end{tabular}
 }
 \label{tab:table_ensemble}
\end{table*}

\subsection{Class Imbalance}
\label{subsec:ablation_class_imbalance}
Class imbalance is a common issue in nuclear instance segmentation, where the majority class (background) dominates the minority class (instances). To address this challenge, we experimented with focal loss and weighted cross-entropy loss on the Lizard dataset with our method. The results, as shown in Table~\ref{tab:table_data_imbalance}, demonstrate that these technique can indeed improve performance. However, we also acknowledge that these techniques often require dataset-spefic tuning, {which can be time-consuming and labor-intensive.} Nevertheless, our method still achieved good performance without the need for these techniques.

\subsection{Cross Validation}
As shown in Table~\ref{tab:table_data_fold}, our cross-validation experiments consistently show that our model performs robustly across different folds, indicating its strong generalization capabilities. The slight variability in results could possibly be attributed to the differences between the training and test datasets in each fold. Fold 2 of Lizard dataset stands out with results that are notably different from the rest, possibly due to unique characteristics or biases present in that specific subset. Despite this outlier, the overall consistency across the other folds confirms the reliability and generalization ability of our model.

\subsection{{Model Regularization}}
\label{subsec:ablation_model_regularization}
{By avoiding overfitting to specific, possibly spurious correlations within the training set, the model with dropout demonstrates improved robustness and accuracy when confronted with the variability inherent in real-world data. Our experimental results, as evidenced in Table~\ref{tab:table_regularization}, illustrate the improvement in model performance when dropout is employed as a regularization technique.}

\subsection{Model Ensembling}
\label{subsec:ablation_model_ensembling}
Ensembling combines multiple diverse models to improve robustness. We use two ensemble strategies: 1) \textit{intra-model}: averaging predictions from flips and rotations of the same input, and 2) \textit{inter-model}: averaging predictions from different encoder backbones. {Table~\ref{tab:table_ensemble}} shows both provide consistent gains. The \textit{intra-ensemble} leverages model smoothness over input variations, while the \textit{inter-ensemble} benefits from complementary information in different backbones. Since real-world data varies, ensembling makes our framework adaptable.

In summary, our ablation analyses provide insight into optimal model architecture choices and training strategies for robust histology image analysis. The findings guide the design of our proposed multi-task framework.

{\section{Discussion}}
\label{sec:discussion}
{
\subsection{Practical Usage}
\label{subsec:practical_usage}
Our proposed method has the potential to be used in various practical applications in digital pathology. One potential application is in estimating tumor burden by counting the number of tumor nuclei and dividing the number by the total number of nuclei. This could be a valuable tool for clinicians, as it could aid in prognostic estimation and patient treatment planning for neo-adjuvant therapy. Our model could be embedded in a software that enables pathologists to view whole-slide images with our segmentation overlay, similar to Figure~\ref{fig:results_vis}, and our cellular composition prediction results displayed alongside. This could potentially aid doctors in their diagnostic workflow and ultimately improve patient care.}

\subsection{Limitations}
\label{subsec:limitations}
{A notable constraint of our approach is the computational complexity in our method's ensembling strategy. The necessity to process data across various base models in multiple iterations can lead to increased processing times, which may be considered a drawback for applications demanding real-time analytical capabilities. The computational intensity is primarily due to the extensive parameterization of our model, with a total of 88,739,206 parameters. This parameter count consists of the two encoder models: 33,671,843 parameters for the SEResNeXt50-based model and 55,067,363 parameters for the SEResNeXt101-based model. In comparison, the original HoVerNet model comprises a more modest 37,639,691 parameters.}

{Furthermore, the inference time of our method, measured on a NVidia 3090 GPU, stands at 0.1421 seconds per image. This is notably higher than the 0.0262 seconds required by HoVerNet under identical conditions. The increased inference time can be attributed to the tripled model inference demands for each intra-model ensemble, compounded by the inter-model ensemble, which itself requires two distinct intra-model ensembles. While this imposes certain limitations on the speed of our method, it is a reasonable trade-off for the enhanced accuracy and robustness that our ensembling strategy provides.}

\section{Conclusion}
In this study, we present an enhanced framework for nuclear instance segmentation, classification, and composition prediction in histology images. {Our approach builds upon the popular HoVer-Net architecture and incorporates several enhancements to improve its performance.} Specifically, we employed advanced SEResNeXt encoders to extract richer features, {modified} the decoder for improved efficiency, and {added} dropout for regularization. {Furthermore, we} implemented a two-level model ensembling strategy {to enhance} generalization capabilities. Extensive experiments demonstrate {the effectiveness of these contributions, resulting in significant complementary performance improvements that} enable our framework to surpass state-of-the-art methods on public benchmarks. {Our approach can be effortlessly extended to} nuclear counting and composition regression tasks, achieving top results on the task. Ablation studies provide insights into optimal architectures and training strategies for robust multi-task histology image analysis. The proposed framework has the potential to facilitate accurate cellular characterization and analysis in various pathology applications. Future work involves expanding the framework to additional modalities and pathology applications requiring accurate cellular characterization. To facilitate further research in this direction, we have released our code and models. {Overall, this study demonstrates the effectiveness of our enhanced framework for nuclear instance segmentation,} classification, and composition prediction in histology images, and its potential applications in computational pathology.

\section*{Declaration of generative AI and AI-assisted technologies in the writing process}
During the preparation of this work, the author(s) used \href{https://poe.com/Claude-instant}{Claude-instant} in order to polish the language. After using this tool/service, the author(s) reviewed and edited the content as needed and take(s) full responsibility for the content of the publication.

\section*{Acknowledgement}
This work received funding from National Natural Science Foundation of China (NSFC, 82303419). We acknowledge the support of Shanghai Technical Service Computing  Center of Science and Engineering, Shanghai University.

%%Harvard
\bibliographystyle{model2-names.bst}\biboptions{authoryear}
\bibliography{refs}

\begin{thebibliography}{67}
\expandafter\ifx\csname natexlab\endcsname\relax\def\natexlab#1{#1}\fi
\providecommand{\url}[1]{\texttt{#1}}
\providecommand{\href}[2]{#2}
\providecommand{\path}[1]{#1}
\providecommand{\DOIprefix}{doi:}
\providecommand{\ArXivprefix}{arXiv:}
\providecommand{\URLprefix}{URL: }
\providecommand{\Pubmedprefix}{pmid:}
\providecommand{\doi}[1]{\href{http://dx.doi.org/#1}{\path{#1}}}
\providecommand{\Pubmed}[1]{\href{pmid:#1}{\path{#1}}}
\providecommand{\bibinfo}[2]{#2}
\ifx\xfnm\relax \def\xfnm[#1]{\unskip,\space#1}\fi
%Type = Article
\bibitem[{Abdel-Nasser et~al.(2022)Abdel-Nasser, Singh and
  Mohamed}]{abdel2022efficient}
\bibinfo{author}{Abdel-Nasser, M.}, \bibinfo{author}{Singh, V.K.},
  \bibinfo{author}{Mohamed, E.M.}, \bibinfo{year}{2022}.
\newblock \bibinfo{title}{Efficient staining-invariant nuclei segmentation
  approach using self-supervised deep contrastive network}.
\newblock \bibinfo{journal}{Diagnostics} \bibinfo{volume}{12},
  \bibinfo{pages}{3024}.
%Type = Inproceedings
\bibitem[{Alsubaie et~al.(2018)Alsubaie, Sirinukunwattana, Raza, Snead and
  Rajpoot}]{alsubaie2018bottom}
\bibinfo{author}{Alsubaie, N.}, \bibinfo{author}{Sirinukunwattana, K.},
  \bibinfo{author}{Raza, S.E.A.}, \bibinfo{author}{Snead, D.},
  \bibinfo{author}{Rajpoot, N.}, \bibinfo{year}{2018}.
\newblock \bibinfo{title}{A bottom-up approach for tumour differentiation in
  whole slide images of lung adenocarcinoma}, in: \bibinfo{booktitle}{Medical
  Imaging 2018: Digital Pathology}, \bibinfo{organization}{International
  Society for Optics and Photonics}. p. \bibinfo{pages}{105810E}.
%Type = Inproceedings
\bibitem[{Bancher et~al.(2021)Bancher, Mahbod, Ellinger, Ecker and
  Dorffner}]{bancher2021improving}
\bibinfo{author}{Bancher, B.}, \bibinfo{author}{Mahbod, A.},
  \bibinfo{author}{Ellinger, I.}, \bibinfo{author}{Ecker, R.},
  \bibinfo{author}{Dorffner, G.}, \bibinfo{year}{2021}.
\newblock \bibinfo{title}{Improving mask r-cnn for nuclei instance segmentation
  in hematoxylin \& eosin-stained histological images}, in:
  \bibinfo{booktitle}{MICCAI Workshop on Computational Pathology},
  \bibinfo{organization}{PMLR}. pp. \bibinfo{pages}{20--35}.
%Type = Article
\bibitem[{Bersanelli(2020)}]{bersanelli2020tumour}
\bibinfo{author}{Bersanelli, M.}, \bibinfo{year}{2020}.
\newblock \bibinfo{title}{Tumour mutational burden as a driver for treatment
  choice in resistant tumours (and beyond)}.
\newblock \bibinfo{journal}{The Lancet Oncology} \bibinfo{volume}{21},
  \bibinfo{pages}{1255--1257}.
%Type = Incollection
\bibitem[{B{\"u}hlmann(2012)}]{buhlmann2012bagging}
\bibinfo{author}{B{\"u}hlmann, P.}, \bibinfo{year}{2012}.
\newblock \bibinfo{title}{Bagging, boosting and ensemble methods}, in:
  \bibinfo{booktitle}{Handbook of computational statistics}.
  \bibinfo{publisher}{Springer}, pp. \bibinfo{pages}{985--1022}.
%Type = Article
\bibitem[{Cai and Vasconcelos(2019)}]{Cai_2019}
\bibinfo{author}{Cai, Z.}, \bibinfo{author}{Vasconcelos, N.},
  \bibinfo{year}{2019}.
\newblock \bibinfo{title}{Cascade r-cnn: High quality object detection and
  instance segmentation}.
\newblock \bibinfo{journal}{IEEE Transactions on Pattern Analysis and Machine
  Intelligence} , \bibinfo{pages}{1–1}\URLprefix
  \url{http://dx.doi.org/10.1109/tpami.2019.2956516},
  \DOIprefix\doi{10.1109/tpami.2019.2956516}.
%Type = Inproceedings
\bibitem[{Cao et~al.(2022)Cao, Wang, Chen, Jiang, Zhang, Tian and
  Wang}]{swinunet}
\bibinfo{author}{Cao, H.}, \bibinfo{author}{Wang, Y.}, \bibinfo{author}{Chen,
  J.}, \bibinfo{author}{Jiang, D.}, \bibinfo{author}{Zhang, X.},
  \bibinfo{author}{Tian, Q.}, \bibinfo{author}{Wang, M.}, \bibinfo{year}{2022}.
\newblock \bibinfo{title}{Swin-unet: Unet-like pure transformer for medical
  image segmentation}, in: \bibinfo{booktitle}{Proceedings of the European
  Conference on Computer Vision Workshops(ECCVW)}.
%Type = Article
\bibitem[{Chan et~al.(1996)Chan, Leung and Wong}]{chan1996expert}
\bibinfo{author}{Chan, S.W.}, \bibinfo{author}{Leung, K.S.},
  \bibinfo{author}{Wong, W.F.}, \bibinfo{year}{1996}.
\newblock \bibinfo{title}{An expert system for the detection of cervical cancer
  cells using knowledge-based image analyzer}.
\newblock \bibinfo{journal}{Artificial Intelligence in Medicine}
  \bibinfo{volume}{8}, \bibinfo{pages}{67--90}.
%Type = Article
\bibitem[{Chen et~al.(2011)Chen, Yang, Liu and Liu}]{chen2011support}
\bibinfo{author}{Chen, H.L.}, \bibinfo{author}{Yang, B.}, \bibinfo{author}{Liu,
  J.}, \bibinfo{author}{Liu, D.Y.}, \bibinfo{year}{2011}.
\newblock \bibinfo{title}{A support vector machine classifier with rough
  set-based feature selection for breast cancer diagnosis}.
\newblock \bibinfo{journal}{Expert systems with applications}
  \bibinfo{volume}{38}, \bibinfo{pages}{9014--9022}.
%Type = Article
\bibitem[{Ciresan et~al.(2012)Ciresan, Giusti, Gambardella and
  Schmidhuber}]{ciresan2012deep}
\bibinfo{author}{Ciresan, D.}, \bibinfo{author}{Giusti, A.},
  \bibinfo{author}{Gambardella, L.}, \bibinfo{author}{Schmidhuber, J.},
  \bibinfo{year}{2012}.
\newblock \bibinfo{title}{Deep neural networks segment neuronal membranes in
  electron microscopy images}.
\newblock \bibinfo{journal}{Advances in neural information processing systems}
  \bibinfo{volume}{25}.
%Type = Inproceedings
\bibitem[{Cire{\c{s}}an et~al.(2013)Cire{\c{s}}an, Giusti, Gambardella and
  Schmidhuber}]{cirecsan2013mitosis}
\bibinfo{author}{Cire{\c{s}}an, D.C.}, \bibinfo{author}{Giusti, A.},
  \bibinfo{author}{Gambardella, L.M.}, \bibinfo{author}{Schmidhuber, J.},
  \bibinfo{year}{2013}.
\newblock \bibinfo{title}{Mitosis detection in breast cancer histology images
  with deep neural networks}, in: \bibinfo{booktitle}{International conference
  on medical image computing and computer-assisted intervention},
  \bibinfo{organization}{Springer}. pp. \bibinfo{pages}{411--418}.
%Type = Article
\bibitem[{Comaniciu et~al.(1999)Comaniciu, Meer and Foran}]{comaniciu1999image}
\bibinfo{author}{Comaniciu, D.}, \bibinfo{author}{Meer, P.},
  \bibinfo{author}{Foran, D.J.}, \bibinfo{year}{1999}.
\newblock \bibinfo{title}{Image-guided decision support system for pathology}.
\newblock \bibinfo{journal}{Machine Vision and Applications}
  \bibinfo{volume}{11}, \bibinfo{pages}{213--224}.
%Type = Inproceedings
\bibitem[{Dawood et~al.(2021)Dawood, Branson, Rajpoot and
  Minhas}]{dawood2021albrt}
\bibinfo{author}{Dawood, M.}, \bibinfo{author}{Branson, K.},
  \bibinfo{author}{Rajpoot, N.M.}, \bibinfo{author}{Minhas, F.},
  \bibinfo{year}{2021}.
\newblock \bibinfo{title}{Albrt: Cellular composition prediction in routine
  histology images}, in: \bibinfo{booktitle}{Proceedings of the IEEE/CVF
  International Conference on Computer Vision}, pp. \bibinfo{pages}{664--673}.
%Type = Article
\bibitem[{Dogar et~al.(2023)Dogar, Shahzad and Fraz}]{dogar2023attention}
\bibinfo{author}{Dogar, G.M.}, \bibinfo{author}{Shahzad, M.},
  \bibinfo{author}{Fraz, M.M.}, \bibinfo{year}{2023}.
\newblock \bibinfo{title}{Attention augmented distance regression and
  classification network for nuclei instance segmentation and type
  classification in histology images}.
\newblock \bibinfo{journal}{Biomedical Signal Processing and Control}
  \bibinfo{volume}{79}, \bibinfo{pages}{104199}.
%Type = Article
\bibitem[{Elmore et~al.(2015)Elmore, Longton, Carney, Geller, Onega, Tosteson,
  Nelson, Pepe, Allison, Schnitt et~al.}]{elmore2015diagnostic}
\bibinfo{author}{Elmore, J.G.}, \bibinfo{author}{Longton, G.M.},
  \bibinfo{author}{Carney, P.A.}, \bibinfo{author}{Geller, B.M.},
  \bibinfo{author}{Onega, T.}, \bibinfo{author}{Tosteson, A.N.},
  \bibinfo{author}{Nelson, H.D.}, \bibinfo{author}{Pepe, M.S.},
  \bibinfo{author}{Allison, K.H.}, \bibinfo{author}{Schnitt, S.J.}, et~al.,
  \bibinfo{year}{2015}.
\newblock \bibinfo{title}{Diagnostic concordance among pathologists
  interpreting breast biopsy specimens}.
\newblock \bibinfo{journal}{Jama} \bibinfo{volume}{313},
  \bibinfo{pages}{1122--1132}.
%Type = Article
\bibitem[{Elston and Ellis(1991)}]{elston1991pathological}
\bibinfo{author}{Elston, C.W.}, \bibinfo{author}{Ellis, I.O.},
  \bibinfo{year}{1991}.
\newblock \bibinfo{title}{Pathological prognostic factors in breast cancer. i.
  the value of histological grade in breast cancer: experience from a large
  study with long-term follow-up}.
\newblock \bibinfo{journal}{Histopathology} \bibinfo{volume}{19},
  \bibinfo{pages}{403--410}.
%Type = Inproceedings
\bibitem[{Fang et~al.(2021)Fang, Yang, Wang, Li, Fang, Shan, Feng and
  Liu}]{Fang_2021_ICCV}
\bibinfo{author}{Fang, Y.}, \bibinfo{author}{Yang, S.}, \bibinfo{author}{Wang,
  X.}, \bibinfo{author}{Li, Y.}, \bibinfo{author}{Fang, C.},
  \bibinfo{author}{Shan, Y.}, \bibinfo{author}{Feng, B.}, \bibinfo{author}{Liu,
  W.}, \bibinfo{year}{2021}.
\newblock \bibinfo{title}{Instances as queries}, in:
  \bibinfo{booktitle}{Proceedings of the IEEE/CVF International Conference on
  Computer Vision (ICCV)}, pp. \bibinfo{pages}{6910--6919}.
%Type = Article
\bibitem[{Fleming et~al.(2012)Fleming, Ravula, Tatishchev and
  Wang}]{fleming2012colorectal}
\bibinfo{author}{Fleming, M.}, \bibinfo{author}{Ravula, S.},
  \bibinfo{author}{Tatishchev, S.F.}, \bibinfo{author}{Wang, H.L.},
  \bibinfo{year}{2012}.
\newblock \bibinfo{title}{Colorectal carcinoma: Pathologic aspects}.
\newblock \bibinfo{journal}{Journal of gastrointestinal oncology}
  \bibinfo{volume}{3}, \bibinfo{pages}{153}.
%Type = Article
\bibitem[{Galli et~al.(2020)Galli, Aguilera, Palermo, Markovic, Nistic{\`o} and
  Signore}]{galli2020relevance}
\bibinfo{author}{Galli, F.}, \bibinfo{author}{Aguilera, J.V.},
  \bibinfo{author}{Palermo, B.}, \bibinfo{author}{Markovic, S.N.},
  \bibinfo{author}{Nistic{\`o}, P.}, \bibinfo{author}{Signore, A.},
  \bibinfo{year}{2020}.
\newblock \bibinfo{title}{Relevance of immune cell and tumor microenvironment
  imaging in the new era of immunotherapy}.
\newblock \bibinfo{journal}{Journal of Experimental \& Clinical Cancer
  Research} \bibinfo{volume}{39}, \bibinfo{pages}{1--21}.
%Type = Inproceedings
\bibitem[{Gamper et~al.(2019)Gamper, Koohbanani, Benet, Khuram and
  Rajpoot}]{gamper2019pannuke}
\bibinfo{author}{Gamper, J.}, \bibinfo{author}{Koohbanani, N.A.},
  \bibinfo{author}{Benet, K.}, \bibinfo{author}{Khuram, A.},
  \bibinfo{author}{Rajpoot, N.}, \bibinfo{year}{2019}.
\newblock \bibinfo{title}{Pannuke: an open pan-cancer histology dataset for
  nuclei instance segmentation and classification}, in:
  \bibinfo{booktitle}{European Congress on Digital Pathology},
  \bibinfo{organization}{Springer}. pp. \bibinfo{pages}{11--19}.
%Type = Article
\bibitem[{Gamper et~al.(2020)Gamper, Koohbanani, Graham, Jahanifar, Khurram,
  Azam, Hewitt and Rajpoot}]{gamper2020pannuke}
\bibinfo{author}{Gamper, J.}, \bibinfo{author}{Koohbanani, N.A.},
  \bibinfo{author}{Graham, S.}, \bibinfo{author}{Jahanifar, M.},
  \bibinfo{author}{Khurram, S.A.}, \bibinfo{author}{Azam, A.},
  \bibinfo{author}{Hewitt, K.}, \bibinfo{author}{Rajpoot, N.},
  \bibinfo{year}{2020}.
\newblock \bibinfo{title}{Pannuke dataset extension, insights and baselines}.
\newblock \bibinfo{journal}{arXiv preprint arXiv:2003.10778} .
%Type = Article
\bibitem[{Graham et~al.(2019a)Graham, Chen, Gamper, Dou, Heng, Snead, Tsang and
  Rajpoot}]{graham2019mild}
\bibinfo{author}{Graham, S.}, \bibinfo{author}{Chen, H.},
  \bibinfo{author}{Gamper, J.}, \bibinfo{author}{Dou, Q.},
  \bibinfo{author}{Heng, P.A.}, \bibinfo{author}{Snead, D.},
  \bibinfo{author}{Tsang, Y.W.}, \bibinfo{author}{Rajpoot, N.},
  \bibinfo{year}{2019}a.
\newblock \bibinfo{title}{Mild-net: Minimal information loss dilated network
  for gland instance segmentation in colon histology images}.
\newblock \bibinfo{journal}{Medical image analysis} \bibinfo{volume}{52},
  \bibinfo{pages}{199--211}.
%Type = Inproceedings
\bibitem[{Graham et~al.(2021a)Graham, Jahanifar, Azam, Nimir, Tsang, Dodd,
  Hero, Sahota, Tank, Benes et~al.}]{graham2021lizard}
\bibinfo{author}{Graham, S.}, \bibinfo{author}{Jahanifar, M.},
  \bibinfo{author}{Azam, A.}, \bibinfo{author}{Nimir, M.},
  \bibinfo{author}{Tsang, Y.W.}, \bibinfo{author}{Dodd, K.},
  \bibinfo{author}{Hero, E.}, \bibinfo{author}{Sahota, H.},
  \bibinfo{author}{Tank, A.}, \bibinfo{author}{Benes, K.}, et~al.,
  \bibinfo{year}{2021}a.
\newblock \bibinfo{title}{Lizard: a large-scale dataset for colonic nuclear
  instance segmentation and classification}, in:
  \bibinfo{booktitle}{Proceedings of the IEEE/CVF International Conference on
  Computer Vision}, pp. \bibinfo{pages}{684--693}.
%Type = Article
\bibitem[{Graham et~al.(2021b)Graham, Jahanifar, Vu, Hadjigeorghiou, Leech,
  Snead, Raza, Minhas and Rajpoot}]{graham2021conic}
\bibinfo{author}{Graham, S.}, \bibinfo{author}{Jahanifar, M.},
  \bibinfo{author}{Vu, Q.D.}, \bibinfo{author}{Hadjigeorghiou, G.},
  \bibinfo{author}{Leech, T.}, \bibinfo{author}{Snead, D.},
  \bibinfo{author}{Raza, S.E.A.}, \bibinfo{author}{Minhas, F.},
  \bibinfo{author}{Rajpoot, N.}, \bibinfo{year}{2021}b.
\newblock \bibinfo{title}{Conic: Colon nuclei identification and counting
  challenge 2022}.
\newblock \bibinfo{journal}{arXiv preprint arXiv:2111.14485} .
%Type = Article
\bibitem[{Graham et~al.(2019b)Graham, Vu, Raza, Azam, Tsang, Kwak and
  Rajpoot}]{graham2019hover}
\bibinfo{author}{Graham, S.}, \bibinfo{author}{Vu, Q.D.},
  \bibinfo{author}{Raza, S.E.A.}, \bibinfo{author}{Azam, A.},
  \bibinfo{author}{Tsang, Y.W.}, \bibinfo{author}{Kwak, J.T.},
  \bibinfo{author}{Rajpoot, N.}, \bibinfo{year}{2019}b.
\newblock \bibinfo{title}{Hover-net: Simultaneous segmentation and
  classification of nuclei in multi-tissue histology images}.
\newblock \bibinfo{journal}{Medical Image Analysis} , \bibinfo{pages}{101563}.
%Type = Article
\bibitem[{Haroske et~al.(1996)Haroske, Dimmer, Friedrich, Meyer, Thieme,
  Theissig and Kunze}]{haroske1996nuclear}
\bibinfo{author}{Haroske, G.}, \bibinfo{author}{Dimmer, V.},
  \bibinfo{author}{Friedrich, K.}, \bibinfo{author}{Meyer, W.},
  \bibinfo{author}{Thieme, B.}, \bibinfo{author}{Theissig, F.},
  \bibinfo{author}{Kunze, K.}, \bibinfo{year}{1996}.
\newblock \bibinfo{title}{Nuclear image analysis of immunohistochemically
  stained cells in breast carcinomas}.
\newblock \bibinfo{journal}{Histochemistry and cell biology}
  \bibinfo{volume}{105}, \bibinfo{pages}{479--485}.
%Type = Article
\bibitem[{Hassan et~al.(2021)Hassan, Saleh, Abdel-Nasser, Omer and
  Puig}]{hassan2021efficient}
\bibinfo{author}{Hassan, L.}, \bibinfo{author}{Saleh, A.},
  \bibinfo{author}{Abdel-Nasser, M.}, \bibinfo{author}{Omer, O.A.},
  \bibinfo{author}{Puig, D.}, \bibinfo{year}{2021}.
\newblock \bibinfo{title}{Efficient multi-organ multi-center cell nuclei
  segmentation method based on deep learnable aggregation network.}
\newblock \bibinfo{journal}{Traitement du Signal} \bibinfo{volume}{38}.
%Type = Inproceedings
\bibitem[{He et~al.(2017)He, Gkioxari, Doll{\'a}r and Girshick}]{he2017mask}
\bibinfo{author}{He, K.}, \bibinfo{author}{Gkioxari, G.},
  \bibinfo{author}{Doll{\'a}r, P.}, \bibinfo{author}{Girshick, R.},
  \bibinfo{year}{2017}.
\newblock \bibinfo{title}{Mask r-cnn}, in: \bibinfo{booktitle}{Proceedings of
  the IEEE international conference on computer vision}, pp.
  \bibinfo{pages}{2961--2969}.
%Type = Inproceedings
\bibitem[{He et~al.(2016)He, Zhang, Ren and Sun}]{he2016deep}
\bibinfo{author}{He, K.}, \bibinfo{author}{Zhang, X.}, \bibinfo{author}{Ren,
  S.}, \bibinfo{author}{Sun, J.}, \bibinfo{year}{2016}.
\newblock \bibinfo{title}{Deep residual learning for image recognition}, in:
  \bibinfo{booktitle}{Proceedings of the IEEE conference on computer vision and
  pattern recognition}, pp. \bibinfo{pages}{770--778}.
%Type = Article
\bibitem[{He et~al.(2021)He, Minn, Solnica-Krezel, Anastasio and
  Li}]{he2021deeply}
\bibinfo{author}{He, S.}, \bibinfo{author}{Minn, K.T.},
  \bibinfo{author}{Solnica-Krezel, L.}, \bibinfo{author}{Anastasio, M.A.},
  \bibinfo{author}{Li, H.}, \bibinfo{year}{2021}.
\newblock \bibinfo{title}{Deeply-supervised density regression for automatic
  cell counting in microscopy images}.
\newblock \bibinfo{journal}{Medical Image Analysis} \bibinfo{volume}{68},
  \bibinfo{pages}{101892}.
%Type = Article
\bibitem[{H{\"o}rst et~al.(2024)H{\"o}rst, Rempe, Heine, Seibold, Keyl,
  Baldini, Ugurel, Siveke, Gr{\"u}nwald, Egger et~al.}]{horst2024cellvit}
\bibinfo{author}{H{\"o}rst, F.}, \bibinfo{author}{Rempe, M.},
  \bibinfo{author}{Heine, L.}, \bibinfo{author}{Seibold, C.},
  \bibinfo{author}{Keyl, J.}, \bibinfo{author}{Baldini, G.},
  \bibinfo{author}{Ugurel, S.}, \bibinfo{author}{Siveke, J.},
  \bibinfo{author}{Gr{\"u}nwald, B.}, \bibinfo{author}{Egger, J.}, et~al.,
  \bibinfo{year}{2024}.
\newblock \bibinfo{title}{Cellvit: Vision transformers for precise cell
  segmentation and classification}.
\newblock \bibinfo{journal}{Medical Image Analysis} \bibinfo{volume}{94},
  \bibinfo{pages}{103143}.
%Type = Misc
\bibitem[{Hu et~al.(2019)Hu, Shen, Albanie, Sun and
  Wu}]{hu2019squeezeandexcitation}
\bibinfo{author}{Hu, J.}, \bibinfo{author}{Shen, L.}, \bibinfo{author}{Albanie,
  S.}, \bibinfo{author}{Sun, G.}, \bibinfo{author}{Wu, E.},
  \bibinfo{year}{2019}.
\newblock \bibinfo{title}{Squeeze-and-excitation networks}.
\newblock \href{http://arxiv.org/abs/1709.01507}{\tt arXiv:1709.01507}.
%Type = Inproceedings
\bibitem[{Hu et~al.(2018)Hu, Shen and Sun}]{hu2018squeeze}
\bibinfo{author}{Hu, J.}, \bibinfo{author}{Shen, L.}, \bibinfo{author}{Sun,
  G.}, \bibinfo{year}{2018}.
\newblock \bibinfo{title}{Squeeze-and-excitation networks}, in:
  \bibinfo{booktitle}{Proceedings of the IEEE conference on computer vision and
  pattern recognition}, pp. \bibinfo{pages}{7132--7141}.
%Type = Article
\bibitem[{Irshad et~al.(2013)Irshad, Veillard, Roux and
  Racoceanu}]{irshad2013methods}
\bibinfo{author}{Irshad, H.}, \bibinfo{author}{Veillard, A.},
  \bibinfo{author}{Roux, L.}, \bibinfo{author}{Racoceanu, D.},
  \bibinfo{year}{2013}.
\newblock \bibinfo{title}{Methods for nuclei detection, segmentation, and
  classification in digital histopathology: a review—current status and
  future potential}.
\newblock \bibinfo{journal}{IEEE reviews in biomedical engineering}
  \bibinfo{volume}{7}, \bibinfo{pages}{97--114}.
%Type = Article
\bibitem[{Isensee et~al.(2024)Isensee, Wald, Ulrich, Baumgartner, Roy,
  Maier-Hein and Jaeger}]{isensee2024nnu}
\bibinfo{author}{Isensee, F.}, \bibinfo{author}{Wald, T.},
  \bibinfo{author}{Ulrich, C.}, \bibinfo{author}{Baumgartner, M.},
  \bibinfo{author}{Roy, S.}, \bibinfo{author}{Maier-Hein, K.},
  \bibinfo{author}{Jaeger, P.F.}, \bibinfo{year}{2024}.
\newblock \bibinfo{title}{nnu-net revisited: A call for rigorous validation in
  3d medical image segmentation}.
\newblock \bibinfo{journal}{arXiv preprint arXiv:2404.09556} .
%Type = Article
\bibitem[{Jia et~al.(2009)Jia, Zhu, Ma, Cao, Li and Chen}]{jia2009mechanisms}
\bibinfo{author}{Jia, J.}, \bibinfo{author}{Zhu, F.}, \bibinfo{author}{Ma, X.},
  \bibinfo{author}{Cao, Z.W.}, \bibinfo{author}{Li, Y.X.},
  \bibinfo{author}{Chen, Y.Z.}, \bibinfo{year}{2009}.
\newblock \bibinfo{title}{Mechanisms of drug combinations: interaction and
  network perspectives}.
\newblock \bibinfo{journal}{Nature reviews Drug discovery} \bibinfo{volume}{8},
  \bibinfo{pages}{111--128}.
%Type = Inproceedings
\bibitem[{Khan et~al.(2016)Khan, Gould and Salzmann}]{khan2016deep}
\bibinfo{author}{Khan, A.}, \bibinfo{author}{Gould, S.},
  \bibinfo{author}{Salzmann, M.}, \bibinfo{year}{2016}.
\newblock \bibinfo{title}{Deep convolutional neural networks for human
  embryonic cell counting}, in: \bibinfo{booktitle}{European conference on
  computer vision}, \bibinfo{organization}{Springer}. pp.
  \bibinfo{pages}{339--348}.
%Type = Article
\bibitem[{Kingma and Ba(2014)}]{kingma2014adam}
\bibinfo{author}{Kingma, D.P.}, \bibinfo{author}{Ba, J.}, \bibinfo{year}{2014}.
\newblock \bibinfo{title}{Adam: A method for stochastic optimization}.
\newblock \bibinfo{journal}{arXiv preprint arXiv:1412.6980} .
%Type = Inproceedings
\bibitem[{Kirillov et~al.(2019)Kirillov, He, Girshick, Rother and
  Doll{\'a}r}]{kirillov2019panoptic}
\bibinfo{author}{Kirillov, A.}, \bibinfo{author}{He, K.},
  \bibinfo{author}{Girshick, R.}, \bibinfo{author}{Rother, C.},
  \bibinfo{author}{Doll{\'a}r, P.}, \bibinfo{year}{2019}.
\newblock \bibinfo{title}{Panoptic segmentation}, in:
  \bibinfo{booktitle}{Proceedings of the IEEE/CVF Conference on Computer Vision
  and Pattern Recognition}, pp. \bibinfo{pages}{9404--9413}.
%Type = Inproceedings
\bibitem[{Kirillov et~al.(2023)Kirillov, Mintun, Ravi, Mao, Rolland, Gustafson,
  Xiao, Whitehead, Berg, Lo et~al.}]{kirillov2023segment}
\bibinfo{author}{Kirillov, A.}, \bibinfo{author}{Mintun, E.},
  \bibinfo{author}{Ravi, N.}, \bibinfo{author}{Mao, H.},
  \bibinfo{author}{Rolland, C.}, \bibinfo{author}{Gustafson, L.},
  \bibinfo{author}{Xiao, T.}, \bibinfo{author}{Whitehead, S.},
  \bibinfo{author}{Berg, A.C.}, \bibinfo{author}{Lo, W.Y.}, et~al.,
  \bibinfo{year}{2023}.
\newblock \bibinfo{title}{Segment anything}, in:
  \bibinfo{booktitle}{Proceedings of the IEEE/CVF International Conference on
  Computer Vision}, pp. \bibinfo{pages}{4015--4026}.
%Type = Article
\bibitem[{Kotu and Deshpande(2015)}]{kotu2015data}
\bibinfo{author}{Kotu, V.}, \bibinfo{author}{Deshpande, B.},
  \bibinfo{year}{2015}.
\newblock \bibinfo{title}{Data mining process}.
\newblock \bibinfo{journal}{Predictive analytics and data mining}
  \bibinfo{volume}{1}, \bibinfo{pages}{17--36}.
%Type = Book
\bibitem[{Kotu and Deshpande(2018)}]{kotu2018data}
\bibinfo{author}{Kotu, V.}, \bibinfo{author}{Deshpande, B.},
  \bibinfo{year}{2018}.
\newblock \bibinfo{title}{Data science: concepts and practice}.
\newblock \bibinfo{publisher}{Morgan Kaufmann}.
%Type = Article
\bibitem[{Li et~al.(2018)Li, Jiang, Zhang, Wang, Wang, Zheng and
  Menze}]{li2018fully}
\bibinfo{author}{Li, H.}, \bibinfo{author}{Jiang, G.}, \bibinfo{author}{Zhang,
  J.}, \bibinfo{author}{Wang, R.}, \bibinfo{author}{Wang, Z.},
  \bibinfo{author}{Zheng, W.S.}, \bibinfo{author}{Menze, B.},
  \bibinfo{year}{2018}.
\newblock \bibinfo{title}{Fully convolutional network ensembles for white
  matter hyperintensities segmentation in mr images}.
\newblock \bibinfo{journal}{NeuroImage} \bibinfo{volume}{183},
  \bibinfo{pages}{650--665}.
%Type = Article
\bibitem[{Lin et~al.(2014)Lin, Chen, Qiu, Wu, Krishnan and Zou}]{lin2014libd3c}
\bibinfo{author}{Lin, C.}, \bibinfo{author}{Chen, W.}, \bibinfo{author}{Qiu,
  C.}, \bibinfo{author}{Wu, Y.}, \bibinfo{author}{Krishnan, S.},
  \bibinfo{author}{Zou, Q.}, \bibinfo{year}{2014}.
\newblock \bibinfo{title}{Libd3c: ensemble classifiers with a clustering and
  dynamic selection strategy}.
\newblock \bibinfo{journal}{Neurocomputing} \bibinfo{volume}{123},
  \bibinfo{pages}{424--435}.
%Type = Inproceedings
\bibitem[{Liu et~al.(2021)Liu, Lin, Cao, Hu, Wei, Zhang, Lin and
  Guo}]{liu2021Swin}
\bibinfo{author}{Liu, Z.}, \bibinfo{author}{Lin, Y.}, \bibinfo{author}{Cao,
  Y.}, \bibinfo{author}{Hu, H.}, \bibinfo{author}{Wei, Y.},
  \bibinfo{author}{Zhang, Z.}, \bibinfo{author}{Lin, S.}, \bibinfo{author}{Guo,
  B.}, \bibinfo{year}{2021}.
\newblock \bibinfo{title}{Swin transformer: Hierarchical vision transformer
  using shifted windows}, in: \bibinfo{booktitle}{Proceedings of the IEEE/CVF
  International Conference on Computer Vision (ICCV)}.
%Type = Article
\bibitem[{Lu et~al.(2018)Lu, Romo-Bucheli, Wang, Janowczyk, Ganesan, Gilmore,
  Rimm and Madabhushi}]{lu2018nuclear}
\bibinfo{author}{Lu, C.}, \bibinfo{author}{Romo-Bucheli, D.},
  \bibinfo{author}{Wang, X.}, \bibinfo{author}{Janowczyk, A.},
  \bibinfo{author}{Ganesan, S.}, \bibinfo{author}{Gilmore, H.},
  \bibinfo{author}{Rimm, D.}, \bibinfo{author}{Madabhushi, A.},
  \bibinfo{year}{2018}.
\newblock \bibinfo{title}{Nuclear shape and orientation features from h\&e
  images predict survival in early-stage estrogen receptor-positive breast
  cancers}.
\newblock \bibinfo{journal}{Laboratory investigation} \bibinfo{volume}{98},
  \bibinfo{pages}{1438--1448}.
%Type = Article
\bibitem[{Polikar(2006)}]{polikar2006ensemble}
\bibinfo{author}{Polikar, R.}, \bibinfo{year}{2006}.
\newblock \bibinfo{title}{Ensemble based systems in decision making}.
\newblock \bibinfo{journal}{IEEE Circuits and systems magazine}
  \bibinfo{volume}{6}, \bibinfo{pages}{21--45}.
%Type = Article
\bibitem[{Prabhu et~al.()Prabhu, Bishnoi and Minu}]{prabhusegmentation}
\bibinfo{author}{Prabhu, A.V.}, \bibinfo{author}{Bishnoi, J.},
  \bibinfo{author}{Minu, R.}, .
\newblock \bibinfo{title}{Segmentation of neoplastic cell nuclei for assisted
  cell labelling using mask r-cnn} .
%Type = Article
\bibitem[{Rubin et~al.(2008)Rubin, Strayer, Rubin and
  Pathology}]{rubin2008clinicopathologic}
\bibinfo{author}{Rubin, R.}, \bibinfo{author}{Strayer, D.},
  \bibinfo{author}{Rubin, E.}, \bibinfo{author}{Pathology, R.},
  \bibinfo{year}{2008}.
\newblock \bibinfo{title}{Clinicopathologic foundations of medicine}.
\newblock \bibinfo{journal}{Lippincot Williams \& Wilkins} .
%Type = Inproceedings
\bibitem[{Schmidt et~al.(2018)Schmidt, Weigert, Broaddus and
  Myers}]{schmidt2018}
\bibinfo{author}{Schmidt, U.}, \bibinfo{author}{Weigert, M.},
  \bibinfo{author}{Broaddus, C.}, \bibinfo{author}{Myers, G.},
  \bibinfo{year}{2018}.
\newblock \bibinfo{title}{Cell detection with star-convex polygons}, in:
  \bibinfo{booktitle}{Medical Image Computing and Computer Assisted
  Intervention - {MICCAI} 2018 - 21st International Conference, Granada, Spain,
  September 16-20, 2018, Proceedings, Part {II}}, pp.
  \bibinfo{pages}{265--273}.
\newblock \DOIprefix\doi{10.1007/978-3-030-00934-2_30}.
%Type = Article
\bibitem[{Sedaghat-Hamedani et~al.(2017)Sedaghat-Hamedani, Haas, Zhu, Geier,
  Kayvanpour, Liss, Lai, Frese, Pribe-Wolferts, Amr
  et~al.}]{sedaghat2017clinical}
\bibinfo{author}{Sedaghat-Hamedani, F.}, \bibinfo{author}{Haas, J.},
  \bibinfo{author}{Zhu, F.}, \bibinfo{author}{Geier, C.},
  \bibinfo{author}{Kayvanpour, E.}, \bibinfo{author}{Liss, M.},
  \bibinfo{author}{Lai, A.}, \bibinfo{author}{Frese, K.},
  \bibinfo{author}{Pribe-Wolferts, R.}, \bibinfo{author}{Amr, A.}, et~al.,
  \bibinfo{year}{2017}.
\newblock \bibinfo{title}{Clinical genetics and outcome of left ventricular
  non-compaction cardiomyopathy}.
\newblock \bibinfo{journal}{European heart journal} \bibinfo{volume}{38},
  \bibinfo{pages}{3449--3460}.
%Type = Article
\bibitem[{Shen et~al.(2016)Shen, Chen, Yu, Kang, Zhang, Li, Yang and
  Liu}]{shen2016evolving}
\bibinfo{author}{Shen, L.}, \bibinfo{author}{Chen, H.}, \bibinfo{author}{Yu,
  Z.}, \bibinfo{author}{Kang, W.}, \bibinfo{author}{Zhang, B.},
  \bibinfo{author}{Li, H.}, \bibinfo{author}{Yang, B.}, \bibinfo{author}{Liu,
  D.}, \bibinfo{year}{2016}.
\newblock \bibinfo{title}{Evolving support vector machines using fruit fly
  optimization for medical data classification}.
\newblock \bibinfo{journal}{Knowledge-Based Systems} \bibinfo{volume}{96},
  \bibinfo{pages}{61--75}.
%Type = Article
\bibitem[{Sirinukunwattana et~al.(2017)Sirinukunwattana, Pluim, Chen, Qi, Heng,
  Guo, Wang, Matuszewski, Bruni, Sanchez et~al.}]{sirinukunwattana2017gland}
\bibinfo{author}{Sirinukunwattana, K.}, \bibinfo{author}{Pluim, J.P.},
  \bibinfo{author}{Chen, H.}, \bibinfo{author}{Qi, X.}, \bibinfo{author}{Heng,
  P.A.}, \bibinfo{author}{Guo, Y.B.}, \bibinfo{author}{Wang, L.Y.},
  \bibinfo{author}{Matuszewski, B.J.}, \bibinfo{author}{Bruni, E.},
  \bibinfo{author}{Sanchez, U.}, et~al., \bibinfo{year}{2017}.
\newblock \bibinfo{title}{Gland segmentation in colon histology images: The
  glas challenge contest}.
\newblock \bibinfo{journal}{Medical image analysis} \bibinfo{volume}{35},
  \bibinfo{pages}{489--502}.
%Type = Article
\bibitem[{Sirinukunwattana et~al.(2016)Sirinukunwattana, Raza, Tsang, Snead,
  Cree and Rajpoot}]{sirinukunwattana2016locality}
\bibinfo{author}{Sirinukunwattana, K.}, \bibinfo{author}{Raza, S.E.A.},
  \bibinfo{author}{Tsang, Y.W.}, \bibinfo{author}{Snead, D.R.},
  \bibinfo{author}{Cree, I.A.}, \bibinfo{author}{Rajpoot, N.M.},
  \bibinfo{year}{2016}.
\newblock \bibinfo{title}{Locality sensitive deep learning for detection and
  classification of nuclei in routine colon cancer histology images}.
\newblock \bibinfo{journal}{IEEE transactions on medical imaging}
  \bibinfo{volume}{35}, \bibinfo{pages}{1196--1206}.
%Type = Article
\bibitem[{Sirinukunwattana et~al.(2018)Sirinukunwattana, Snead, Epstein, Aftab,
  Mujeeb, Tsang, Cree and Rajpoot}]{sirinukunwattana2018novel}
\bibinfo{author}{Sirinukunwattana, K.}, \bibinfo{author}{Snead, D.},
  \bibinfo{author}{Epstein, D.}, \bibinfo{author}{Aftab, Z.},
  \bibinfo{author}{Mujeeb, I.}, \bibinfo{author}{Tsang, Y.W.},
  \bibinfo{author}{Cree, I.}, \bibinfo{author}{Rajpoot, N.},
  \bibinfo{year}{2018}.
\newblock \bibinfo{title}{Novel digital signatures of tissue phenotypes for
  predicting distant metastasis in colorectal cancer}.
\newblock \bibinfo{journal}{Scientific reports} \bibinfo{volume}{8},
  \bibinfo{pages}{1--13}.
%Type = Article
\bibitem[{Srinidhi et~al.(2020)Srinidhi, Ciga and Martel}]{srinidhi2020deep}
\bibinfo{author}{Srinidhi, C.L.}, \bibinfo{author}{Ciga, O.},
  \bibinfo{author}{Martel, A.L.}, \bibinfo{year}{2020}.
\newblock \bibinfo{title}{Deep neural network models for computational
  histopathology: A survey}.
\newblock \bibinfo{journal}{Medical Image Analysis} , \bibinfo{pages}{101813}.
%Type = Article
\bibitem[{Srinidhi et~al.(2021)Srinidhi, Ciga and Martel}]{srinidhi2021deep}
\bibinfo{author}{Srinidhi, C.L.}, \bibinfo{author}{Ciga, O.},
  \bibinfo{author}{Martel, A.L.}, \bibinfo{year}{2021}.
\newblock \bibinfo{title}{Deep neural network models for computational
  histopathology: A survey}.
\newblock \bibinfo{journal}{Medical Image Analysis} \bibinfo{volume}{67},
  \bibinfo{pages}{101813}.
%Type = Article
\bibitem[{Verma et~al.(2021)Verma, Kumar, Patil, Kurian, Rane, Graham, Vu,
  Zwager, Raza, Rajpoot et~al.}]{verma2021monusac2020}
\bibinfo{author}{Verma, R.}, \bibinfo{author}{Kumar, N.},
  \bibinfo{author}{Patil, A.}, \bibinfo{author}{Kurian, N.C.},
  \bibinfo{author}{Rane, S.}, \bibinfo{author}{Graham, S.},
  \bibinfo{author}{Vu, Q.D.}, \bibinfo{author}{Zwager, M.},
  \bibinfo{author}{Raza, S.E.A.}, \bibinfo{author}{Rajpoot, N.}, et~al.,
  \bibinfo{year}{2021}.
\newblock \bibinfo{title}{Monusac2020: A multi-organ nuclei segmentation and
  classification challenge}.
\newblock \bibinfo{journal}{IEEE Transactions on Medical Imaging}
  \bibinfo{volume}{40}, \bibinfo{pages}{3413--3423}.
%Type = Article
\bibitem[{Vo and Kim(2023)}]{vo2023mulvernet}
\bibinfo{author}{Vo, V.T.T.}, \bibinfo{author}{Kim, S.H.},
  \bibinfo{year}{2023}.
\newblock \bibinfo{title}{Mulvernet: nucleus segmentation and classification of
  pathology images using the hover-net and multiple filter units}.
\newblock \bibinfo{journal}{Electronics} \bibinfo{volume}{12},
  \bibinfo{pages}{355}.
%Type = Article
\bibitem[{Wang et~al.(2023)Wang, Qin, Chen, Wang, Han, Zhu and
  Qiao}]{wang2023improved}
\bibinfo{author}{Wang, J.}, \bibinfo{author}{Qin, L.}, \bibinfo{author}{Chen,
  D.}, \bibinfo{author}{Wang, J.}, \bibinfo{author}{Han, B.W.},
  \bibinfo{author}{Zhu, Z.}, \bibinfo{author}{Qiao, G.}, \bibinfo{year}{2023}.
\newblock \bibinfo{title}{An improved hover-net for nuclear segmentation and
  classification in histopathology images}.
\newblock \bibinfo{journal}{Neural Computing and Applications}
  \bibinfo{volume}{35}, \bibinfo{pages}{14403--14417}.
%Type = Inproceedings
\bibitem[{Weigert and Schmidt(2022)}]{weigert2022}
\bibinfo{author}{Weigert, M.}, \bibinfo{author}{Schmidt, U.},
  \bibinfo{year}{2022}.
\newblock \bibinfo{title}{Nuclei instance segmentation and classification in
  histopathology images with stardist}, in: \bibinfo{booktitle}{The IEEE
  International Symposium on Biomedical Imaging Challenges (ISBIC)}.
\newblock \DOIprefix\doi{10.1109/ISBIC56247.2022.9854534}.
%Type = Inproceedings
\bibitem[{Weigert et~al.(2020)Weigert, Schmidt, Haase, Sugawara and
  Myers}]{weigert2020}
\bibinfo{author}{Weigert, M.}, \bibinfo{author}{Schmidt, U.},
  \bibinfo{author}{Haase, R.}, \bibinfo{author}{Sugawara, K.},
  \bibinfo{author}{Myers, G.}, \bibinfo{year}{2020}.
\newblock \bibinfo{title}{Star-convex polyhedra for 3d object detection and
  segmentation in microscopy}, in: \bibinfo{booktitle}{The IEEE Winter
  Conference on Applications of Computer Vision (WACV)}.
\newblock \DOIprefix\doi{10.1109/WACV45572.2020.9093435}.
%Type = Article
\bibitem[{Xing et~al.(2013)Xing, Su, Neltner and Yang}]{xing2013automatic}
\bibinfo{author}{Xing, F.}, \bibinfo{author}{Su, H.}, \bibinfo{author}{Neltner,
  J.}, \bibinfo{author}{Yang, L.}, \bibinfo{year}{2013}.
\newblock \bibinfo{title}{Automatic ki-67 counting using robust cell detection
  and online dictionary learning}.
\newblock \bibinfo{journal}{IEEE Transactions on Biomedical Engineering}
  \bibinfo{volume}{61}, \bibinfo{pages}{859--870}.
%Type = Inproceedings
\bibitem[{Xue et~al.(2016)Xue, Ray, Hugh and Bigras}]{xue2016cell}
\bibinfo{author}{Xue, Y.}, \bibinfo{author}{Ray, N.}, \bibinfo{author}{Hugh,
  J.}, \bibinfo{author}{Bigras, G.}, \bibinfo{year}{2016}.
\newblock \bibinfo{title}{Cell counting by regression using convolutional
  neural network}, in: \bibinfo{booktitle}{European Conference on Computer
  Vision}, \bibinfo{organization}{Springer}. pp. \bibinfo{pages}{274--290}.
%Type = Article
\bibitem[{Yao et~al.(2021)Yao, Huang, Sun, Hussain and Jude}]{yao2021pointnu}
\bibinfo{author}{Yao, K.}, \bibinfo{author}{Huang, K.}, \bibinfo{author}{Sun,
  J.}, \bibinfo{author}{Hussain, A.}, \bibinfo{author}{Jude, C.},
  \bibinfo{year}{2021}.
\newblock \bibinfo{title}{Pointnu-net: Simultaneous multi-tissue histology
  nuclei segmentation and classification in the clinical wild}.
\newblock \bibinfo{journal}{arXiv preprint arXiv:2111.01557} .
%Type = Inproceedings
\bibitem[{Zhu et~al.(2021)Zhu, Chen, Zheng, Zhang and Wang}]{zhu2021real}
\bibinfo{author}{Zhu, Y.}, \bibinfo{author}{Chen, Z.}, \bibinfo{author}{Zheng,
  Y.}, \bibinfo{author}{Zhang, Q.}, \bibinfo{author}{Wang, X.},
  \bibinfo{year}{2021}.
\newblock \bibinfo{title}{Real-time cell counting in unlabeled microscopy
  images}, in: \bibinfo{booktitle}{Proceedings of the IEEE/CVF International
  Conference on Computer Vision}, pp. \bibinfo{pages}{694--703}.
%Type = Article
\bibitem[{Zou et~al.(2018)Zou, Qu, Luo, Yin, Ju and Tang}]{zou2018predicting}
\bibinfo{author}{Zou, Q.}, \bibinfo{author}{Qu, K.}, \bibinfo{author}{Luo, Y.},
  \bibinfo{author}{Yin, D.}, \bibinfo{author}{Ju, Y.}, \bibinfo{author}{Tang,
  H.}, \bibinfo{year}{2018}.
\newblock \bibinfo{title}{Predicting diabetes mellitus with machine learning
  techniques}.
\newblock \bibinfo{journal}{Frontiers in genetics} \bibinfo{volume}{9},
  \bibinfo{pages}{515}.

\end{thebibliography}

\end{document}